\documentclass[aps,notitlepage,superscriptaddress,floatfix,pra,twocolumn]{revtex4-1}

\usepackage{lipsum}
\usepackage{graphicx}
\usepackage{amsmath,amssymb}
\usepackage{braket}
\usepackage{mathtools}
\usepackage{hyperref}
\usepackage{multirow}
\usepackage{color}
\usepackage[normalem]{ulem}
\usepackage{amsfonts}
\usepackage{float}
\usepackage{pdfpages} 
\makeatletter
\usepackage{subfigure}
\usepackage{soul, xcolor}
\usepackage{dsfont}
\usepackage{hyperref}
\hypersetup{
    pdfborder={0 0 0}
}

\setstcolor{red}

\newcommand{\sectitle}[1]{\textbf{#1}.}

\def\beq{\begin{equation}}
\def\eeq{\end{equation}}
\def\bea{\begin{eqnarray}}
\def\eea{\end{eqnarray}}

\usepackage{pdfpages} 
\makeatletter
\AtBeginDocument{\let\LS@rot\@undefined}
\makeatother

\begin{document}

\title{Anomalous diffusion and singular transport from hydrodynamic recoupling}

\author{Ewan McCulloch}
\affiliation{Laboratoire de Physique de l'École Normale Supérieure, CNRS, ENS \& Université PSL; 24 rue Lhomond, 75005 Paris, France}
\affiliation{Department of Electrical and Computer Engineering, Princeton University, Princeton, NJ 08544, USA}

\author{Romain Vasseur}
\affiliation{Department of Theoretical Physics, University of Geneva, 24 quai Ernest-Ansermet, 1211 Geneva, Switzerland}
\affiliation{Geneva Quantum Center, University of Geneva}

\author{Sarang Gopalakrishnan}
\affiliation{Department of Electrical and Computer Engineering, Princeton University, Princeton, NJ 08544, USA}

\begin{abstract}
In charge neutral fluids, such as the Dirac fluid in graphene at the Dirac point, charge transport remains diffusive despite the presence of ballistically propagating sound waves: sound waves ``hydrodynamically decouple'' from the slower charge fluctuations. For quasi-one-dimensional
charge neutral fluids, we show that this convective charge diffusion is \emph{not} smoothly connected to the normal diffusion that arises when momentum conservation is broken by noise (or static impurities). Instead, the charge diffusion constant is a discontinuous function of noise, which (in the weak-noise limit) depends only on the ratio of momentum and energy relaxation rates. In the special limit of momentum-conserving noise (e.g., spatially uniform fluctuations of the Hamiltonian), the diffusion constant diverges in the presence of noise. We describe the resulting superdiffusion in terms of coupled Burgers equations. We present a general mechanism---hydrodynamic \emph{recoupling}---by which weak noise can induce singular changes in transport coefficients. Our results highlight the limits of zero-noise extrapolation for predicting dynamical quantities like diffusion constants.
\end{abstract}

\maketitle

Hydrodynamics dictates how systems relax to equilibrium under generic classical or quantum dynamics~\cite{landau1987fluid,Chaikin_Lubensky_1995,forster2018hydrodynamic, spohn2012large, liu2018lectures,Doyon2022}. The structure of hydrodynamics is strongly constrained by the symmetries of the system; long-lived deviations from equilibrium are due to fluctuations of the associated conserved charge densities. Within hydrodynamics, charge fluctuations generically relax diffusively (if momentum is not conserved) or spread ballistically (if it is). In the momentum-conserving case, charge density fluctuations are carried by ballistically propagating sound waves. It has long been appreciated that discrete symmetries can change this picture: for example, a charge conjugation or particle-hole symmetry can prevent sound waves from coupling to a particular conserved charge~\cite{PhysRevB.57.8307, Fong2012, Lucas2018, anomalousFCS_spinchain_review}. In this setting---which occurs most famously in graphene at charge neutrality~\cite{crossno2016observation,majumdar2025universality}---ballistic energy transport coexists with diffusive charge transport~\footnote{For unscreened long-range Coulomb interactions, charge dynamics in graphene is non-diffusive at long wavelengths and is therefore outside the scope of the present work. In graphene screened by a nearby metallic gate, the interaction becomes effectively short-ranged at long wavelengths, restoring a diffusive charge mode at neutrality~\cite{PhysRevB.103.115402,Kim2020}.}. This charge diffusion has a convective origin and qualitatively differs from the ``regular'' hydrodynamic diffusion that occurs in the presence of momentum relaxation: as we recently showed~\cite{Gopalakrishnan2022a,pnas.2403327121,PhysRevE.111.015410}, in quasi-1D geometries it leads to anomalous noise signatures, such as non-gaussian full counting statistics. (The coexistence of charge diffusion and ballistic modes can, in fact, occur in charge-neutral systems even without a discrete symmetry~\cite{Krajnik2024}.)

\begin{figure*}[t]
\begin{center}
\includegraphics[width = \textwidth]{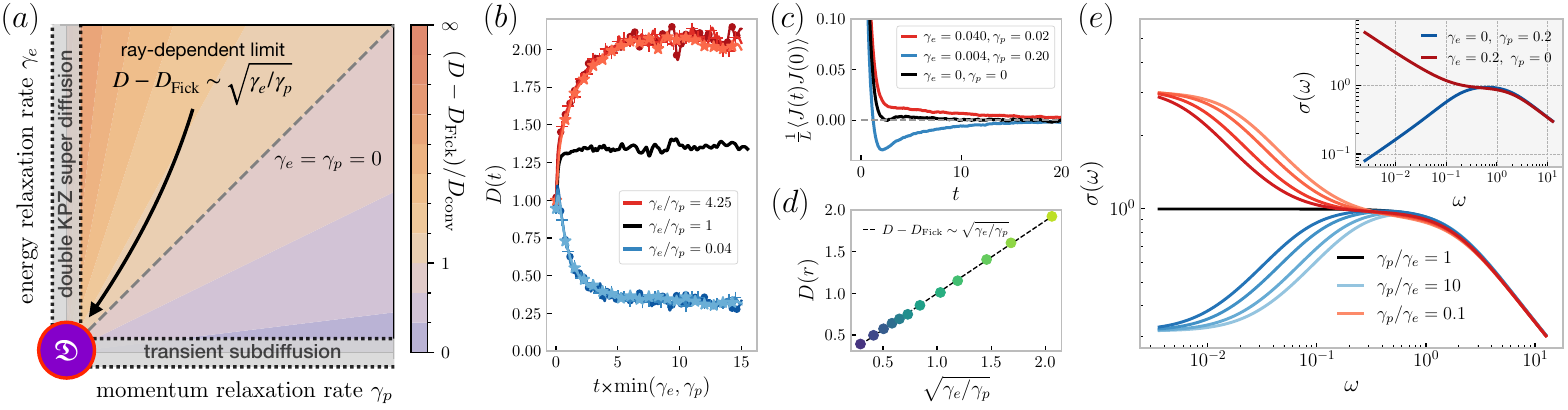}
\caption{\textbf{Singular zero-noise limit of charge diffusion in a particle-hole symmetric fluid}. (a) Schematic of diffusion constant dependence on energy and momentum relaxation rates $\gamma_e,\gamma_p$. The fully conserved point $\mathcal D$ is not reached continuously: approaching it along a fixed $\gamma_e/\gamma_p$ ray gives a ray-dependent diffusion constant (Eq.~\eqref{eq2}) which diverges as $\gamma_p \to 0$. (b) Time-dependent diffusion constant $D(t)$ from equilibrium current fluctuations for several fixed ratios $\gamma_e/\gamma_p$, shown versus $t\times \min(\gamma_e,\gamma_p)$. The black curve is the fully conserved case $\gamma_e=\gamma_p=0$, plotted versus $0.02t$ for visual comparison. For the red (blue) curves from dark to light $\gamma_e=0.006,0.009,0.012$ ($\gamma_e=0.015,0.02,0.03$); $\gamma_p$ is fixed by the ratio stated in the figure. (c) Total-current autocorrelator $L^{-1}\langle J(t)J(0)\rangle$, whose long-time sign (correlated versus anticorrelated) depends on the dominant relaxation channel. (d) Plateau diffusion constant versus $\sqrt{\gamma_e/\gamma_p}$, confirming the predicted square-root dependence. (e) Optical conductivity from fluctuating hydrodynamics, Eq.~\eqref{eq:opticalconductivity}, illustrating the noncommuting zero-frequency and zero-noise limits. For convenience, we set $D_{\rm Fick}=0$ and $c=\chi=\chi_e=\chi_p=D_p=1$. When both energy and momentum conservation are broken, the conductivity has a $\gamma_e/\gamma_p$-dependent zero-frequency limit. Inset: the limiting cases $\gamma_e=0$ and $\gamma_p=0$ give respectively vanishing $(\sigma(\omega)\sim \sqrt{\omega})$ and divergent $(\sigma(\omega)\sim 1/\sqrt{\omega})$ low-frequency convective conductivity. All simulations use system size $L=2000$; the weak-dissipation diffusion constant in (d) is extracted from simulations run to time $t=1000$.
}
\label{fig1}
\end{center}
\end{figure*}

In this work, we explore how this convective charge diffusion crosses over into regular diffusion when one weakly breaks the underlying conservation laws. For simplicity, we first consider a minimal hydrodynamic model with three conserved quantities---which for concreteness we will term energy, momentum, and charge---and a particle-hole symmetry (PHS)~\footnote{For Graphene, this minimal three-mode description assumes that the carrier imbalance is not an additional conserved mode~\cite{PhysRevB.79.085415}. We discuss this further in the Supplemental Material~\cite{suppmat}.}. We also discuss generalizations that do not require a discrete symmetry. Our results are most striking for quasi-one-dimensional wire geometries, so we will focus on these. When all conservation laws are strictly satisfied, charge transport is diffusive, with a diffusion constant $\mathfrak{D}$. We introduce momentum and energy relaxation rates $\gamma_p, \gamma_e$ respectively, and explore how the diffusion constant $D(\gamma_e, \gamma_p)$ evolves as these two relaxation rates are tuned to zero. For generic noise, both relaxation mechanisms will be present, with some nonuniversal ratio. Our main conclusion is that the zero-noise limit is singular (Fig.~\ref{fig1}(a)): in the fully conserved limit, we have
\begin{equation}\label{eq1} \mathfrak{D}\equiv D(\gamma_e=0, \gamma_p=0)= D_{\rm Fick} + D_{\rm conv}. \end{equation}
%
The contribution $D_{\rm Fick}$ is regular and generically nonzero at $\gamma_e=\gamma_p=0$. However, at the fully conserved point, there is an additional convective contribution $D_{\rm conv}$ from the ballistic sound waves
to the charge diffusion constant~\cite{Gopalakrishnan2018,Medenjak2020,pnas.2403327121,PhysRevE.111.015410,PhysRevLett.134.187101}. In the momentum and energy-breaking case, the diffusion constant instead scales as 
\begin{equation}\label{eq2}
D(\gamma_e, \gamma_p)= D_{\rm Fick} + D_{\rm conv} (\gamma_e/\gamma_p)^{1/2} + \cdots,
\end{equation}
with $\cdots$ representing corrections that are analytic in $\gamma_e, \gamma_p$ (and vanishing as $\gamma_{e}, \gamma_{p}\to 0$)~\cite{suppmat}.
This result has important implications for recent computational approaches 
that compute diffusion constants by zero-noise extrapolation~\cite{temme2016error, li2017efficient, RevModPhys.95.045005, PhysRevB.105.245101, xct1-7kf2}: for a broad class of charge-neutral fluids we consider, the diffusion constant approaches a well-defined weak-noise limit that is unrelated to the strictly zero-noise value $\mathfrak D$.
Previous work had established a discontinuity of this type in the fine-tuned case of integrable spin chains~\cite{10.1073/pnas.2202823119} subject to dephasing noise; we show here that this phenomenon is not tied to integrability or to a particular noise model.

Another key result of our work is that the diffusion constant diverges when $\gamma_p = 0$ and $\gamma_e \neq 0$: i.e., when only momentum and charge are conserved. Thus, weak momentum-conserving noise (e.g., global fluctuations of the Hamiltonian parameters) can cause a singular enhancement of charge diffusion, leading to superdiffusive charge transport. The coupled momentum and charge dynamics can be described using a recent hydrodynamic theory~\cite{m_phi_theory, roy2024universality} of coupled Burgers equations. This theory was introduced to describe certain fine-tuned limits of coupled asymmetric exclusion processes; we show how it emerges without fine-tuning. 

\sectitle{Hydrodynamics with weak relaxation}---We consider a one-dimensional fluid with energy density $e$, momentum density $p$, and charge density $n$~\cite{pnas.2403327121,PhysRevE.111.015410,PhysRevE.111.024141,yoshimura2025hydrodynamicfluctuationsstochasticcharged}. Particle-hole symmetry makes $n$ odd under charge conjugation while $e,p$ are even, so charge has no linear overlap with the sound modes. The leading hydrodynamic coupling of charge to sound is therefore nonlinear and convective~\cite{pnas.2403327121}. Including weak relaxation of energy and momentum, the hydrodynamic equations take the form
\bea\label{basichydro}
\partial_t e+\partial_xp&=&-\gamma_e e+\eta_e+\ldots,\\
\partial_t p+c^2\partial_xe&=&-\gamma_p p+D_p\partial_x^2p+\eta_p+\ldots,
\nonumber\\
\partial_t n+\partial_x j_n&=&0,\qquad
j_n=pn-D_{\rm Fick}\partial_x n+\eta_n+\ldots, \nonumber
\eea
where the dissipative coefficients $\gamma_e, \gamma_p, D_p, D_{\rm Fick}$ are accompanied by noise terms $\eta_e, \eta_p, \eta_n$ that obey the fluctuation-dissipation theorem, and where we have normalized $p$ so that the coefficient of $pn$ is one. The equations for $e$, $p$ are simply damped sound modes with speed of sound $c$, while the ellipses include nonlinear corrections to the sound modes (and sub-leading cubic corrections to $j_n$). These corrections broaden the sound peaks sub-ballistically (KPZ in one dimension~\cite{SpohnNLFH,Delacretaz2020}), while the sound modes themselves advect ballistically. Thus the leading effect of sound on charge is its Euler-scale advection and the sound-mode broadening is subleading for the convective contribution computed below~\cite{pnas.2403327121}.

It is convenient to separate the Fickian and convective contributions to charge spreading: $D_{\rm Fick}$ gives ordinary diffusive broadening, while $pn$ advects the charge profile. Since Fickian broadening modifies the convective contribution only at subleading order (as shown below), we set $D_{\rm Fick}=0$ and restore it by convolution at the end~\cite{pnas.2403327121}.
Thus, a charge packet initially near $x_0$ obeys $\dot X=p(X,t)$; since
its motion is sub-ballistic relative to the sound waves, the leading
displacement is
\beq\label{eq:rigid_shift}
X(t)=\int_0^t d\tau\,p(x_0,\tau),
\eeq
where $p(x_0,\tau)$ is obtained by propagating the momentum along characteristics $p(x,t)=p_r(x-ct)+p_l(x+ct)$. (A derivation of this rigid-shift relation is given in the End Matter.) 

The charge diffusion constant is therefore $D_X$, the diffusion constant of $X(t)$; restoring Fickian diffusion gives $D=D_{\rm Fick}+D_X$. We now determine $D$ by the growth of ${\rm Var}[X(t)]$. In Fourier space, this is
\beq\label{fourierform}
\langle X(t)^2 \rangle = \int \frac{dk}{2\pi} \int \frac{d\omega}{2\pi} \frac{\sin^2(\omega t/2)}{(\omega/2)^2} S_{pp}(k, \omega), 
\eeq
where $S_{pp}(k, \omega)$ denotes the Fourier transform of $\langle p(x, t) p(0,0)\rangle$. 
Solving for $p(k,\omega)$ by fourier transforming Eqs.~\eqref{basichydro} (see the supplemental materials~\cite{suppmat} for details) gives
$\langle X^2 \rangle =
2 D(\gamma_e,\gamma_p)\, t$ at long times, with
\begin{equation}\label{mainresult}
    D(\gamma_e,\gamma_p)=\int\frac{dk}{4\pi}\,
\frac{2\gamma_p\chi_p\,\gamma_e^2+2\gamma_e\chi_e\,c^4k^2}
{(\gamma_e\gamma_p+c^2k^2)^2}\sim \sqrt{\frac{\gamma_e}{\gamma_p}},
\end{equation}
where $\chi_e$ and $\chi_p$ are energy and momentum static susceptibilities, and we took $D_p=0$ for simplicity (the broadening of the sound peaks is subleading to the Euler-scale advection in the weak symmetry breaking limit $\gamma_{e},\gamma_p\to 0$).
This expression has two immediate consequences. First, along a general ray in $(\gamma_e, \gamma_p)$ space the diffusion
constant approaches a finite ray-dependent value that does not match continuously to $\mathfrak{D}$ (except along $\gamma_e=\gamma_p$ with $D(\gamma,\gamma)\to D_{\rm conv}$). Second, it suggests that when energy (momentum) remains as a residual conserved quantity, the diffusion constant vanishes (diverges). This approach can be generalized to compute the charge structure factor and optical conductivity~\cite{suppmat}. In the low frequency regime,  fluctuating hydrodynamics predicts
\begin{equation} \label{eq:opticalconductivity}
\sigma(\omega) = \chi \int \frac{dk }{4\pi} S_{pp} (k, \omega).
\end{equation}
In the d.c.~limit, we recover $\sigma = \chi D$ with $D$ given by eq.~\eqref{mainresult}, and where $\chi$ is the charge susceptibility. As illustrated in Fig.~\ref{fig1}, the zero-noise and zero-relaxation rate results do not commute. The discontinuity of the d.c.~conductivity is associated with broad regimes of anomalous  super- or subdiffusive charge transport.  

We can now justify neglecting Fickian diffusion and sound mode broadening when computing the convective contribution in Eq.~\eqref{mainresult}. This
contribution is dominated by fluctuations with wavenumber
$k_*\sim\sqrt{\gamma_e\gamma_p}/c$. Over the sound-mode lifetime
$t_*=\max(\gamma_e^{-1},\gamma_p^{-1})$, Fickian diffusion spreads a
charge element by $\sqrt{D_{\rm Fick}t_*}$, whereas the KPZ broadening of a sound packet grows as
$t_*^{2/3}$. Both are parametrically
smaller than $k_*^{-1}$ as $\gamma_{e,p}\to 0$ at fixed $\gamma_e/\gamma_p$. Thus both Fickian diffusion and sound-mode broadening give only subleading corrections to the charge diffusion constant in the weak-relaxation limit.

\sectitle{Anomalous transport}---To gain an intuition for Eqs.~\eqref{mainresult},~\eqref{eq:opticalconductivity}, we now consider these two limiting cases, for which the analysis simplifies. 
In the energy-conserving limit, a charge packet encountered \emph{anticorrelated} kicks from encountering the same sound wave first moving right, then left. When momentum is conserved and energy is not, one has precisely the opposite effect: a sound wave carrying positive energy (relative to the reference state) backscatters as one carrying negative energy. Both of these waves kick the charge packet in the \emph{same} direction, so charge diffusion is enhanced rather than suppressed. Assuming linear diffusion of momentum, there are ${\cal O}(t^{1/2})$ independent sound waves that cross the origin, each doing so ${\cal O}(t^{1/2})$ times and therefore each contributing a (random signed) kick also of the size ${\cal O}(t^{1/2})$, leading to a total displacement $X(t)\sim t^{1/2}\times t^{1/4}=t^{3/4}$.
This clearly contradicts our assumption of hydrodynamic decoupling (since in the linear theory momentum is diffusive).
To arrive at a consistent theory we need to revisit the hydrodynamic equations for the two remaining conserved quantities, momentum and charge. Including all symmetry-allowed terms, these equations take the form:
\bea\label{coupledburgers}
&& \partial_t p + \partial_x(C_1 p^2 + C_2 n^2) = D_p \partial_x^2 p + \partial_x \eta_p, \nonumber \\
&& \partial_t n + \partial_x (n p) = D_{\rm Fick} \partial_x^2 n + \partial_x \eta_n.
\eea
Eq.~\eqref{coupledburgers} describes a ``degenerate'' limit of two coupled Burgers equations, recently introduced in Ref.~\cite{m_phi_theory} (see also Refs.~\cite{PhysRevLett.69.929,PhysRevE.64.021402}) and explored further in Refs.~\cite{roy2024universality,roy2025fixedpointsuniversalityclasses} 
(where the equations take ``cyclic" form yielding a Gaussian stationary state~\cite{2026arXiv260722527K}). Previously considered realizations of Eq.~\eqref{coupledburgers} required fine-tuning to achieve the degenerate and cyclic conditions; in the present case, these conditions follow from the symmetries of the problem~\footnote{Symmetry constrains the allowed nonlinear couplings but does not require every symmetry-allowed coefficient to be nonzero; accidental zeros can occur in fine-tuned microscopic models~\cite{Alba2026,2026arXiv260728255B}.}. Momentum and charge are strongly nonlinearly coupled, and both quantities are expected to display Kardar-Parisi-Zhang (KPZ)~\cite{PhysRevLett.56.889} scaling $x \sim t^{2/3}$.


\sectitle{Crossover timescales}---We now turn to the crossover timescales in the general case where $\gamma_p, \gamma_e > 0$. To discuss these timescales, it is convenient to invoke the Kubo formula that relates the d.c. conductivity (or equivalently the diffusion constant) to the autocorrelation function of the charge current, $D \propto L^{-1} \int_0^\infty dt\, \langle J_n(t) J_n(0) \rangle$, where $L$ is the system size and $J_n = \int dx \, j_n(x)$ [cf. Eq.~\eqref{basichydro}]. Neglecting the Fickian term, and assuming hydrodynamic decoupling, we have $\langle j_n(x,t) j_n(0,0) \rangle \simeq \langle p(x,t) p(0,0) \rangle \langle n(x,t) n(0,0) \rangle$. Since charge density is parametrically slower than momentum, the charge autocorrelation function is approximately a delta function, so $D \propto \int_0^\infty dt \langle p(0,t) p(0,0)\rangle$. 

In the absence of relaxation processes, $\langle p(x,t) p(x,0)\rangle$ decorrelates rapidly, since sound waves move ballistically. Energy relaxation without momentum relaxation causes positive correlations,
$\langle p(x,t)p(0,0)\rangle \sim (\gamma_e/t)^{1/2}$, for
$t \gtrsim \gamma_e^{-1}$, since momentum is then diffusive with a large return
probability (before crossing over to the strong-coupling regime of Eq.~\eqref{coupledburgers} at long times). Conversely, momentum relaxation without energy relaxation causes sound waves to back-scatter, giving rise to anticorrelations in $\langle p(x,t) p(x,0)\rangle \sim -\gamma_p^{-1/2} t^{-3/2}$ (since a returning sound wave is carrying the same energy but opposite momentum)~\cite{PhysRevB.73.035113}. The slower of the two relaxation processes acts as a long-time cutoff: beyond that timescale, sound waves cease to be defined~\cite{suppmat}, and $n$ is the only surviving hydrodynamic mode. For example, when energy relaxation is faster, the power-law dependence of $\langle p(x,t) p(0,0) \rangle$ is cut off on a timescale $\sim \gamma_p^{-1}$, so the net contribution to the diffusion constant is $\sqrt{\gamma_e/\gamma_p}$. 
The emergent long timescale therefore produces a singular correction to diffusion (Fig.~\ref{fig1}).
We dub this phenomenon ``hydrodynamic recoupling'', as the sound waves that decoupled in the $\gamma_e = \gamma_p=0$ limit ``recouple'' for  $\gamma_e, \gamma_p >0$, and act as a non-Markovian bath.

%

On the diagonal ray $\gamma_e=\gamma_p$ backscattering of sound waves is equally likely to reverse the wave's momentum as not, so that consecutive collisions with a charge element provide uncorrelated kicks. Therefore, charge diffusion in the weak dissipation limit along this special ray corresponds to the fully conserved case---in both cases the kicks contributing to $X(t)$ are effectively Markovian~\cite{suppmat}---fixing the coefficient of $\sqrt{\gamma_e/\gamma_p}$ to $D_{\rm conv}$ in Eq.~\eqref{eq2}. Finally we remark that the scaling form~\eqref{mainresult} will be modified for $\gamma_p/\gamma_e \ll 1$, because the nonlinearities in Eq.~\eqref{coupledburgers} must be included in the solution of Eq.~\eqref{basichydro}. 


\begin{figure}
    \centering
    \includegraphics[width=1.\columnwidth, trim=4mm 4mm 3mm 0mm,
  clip]{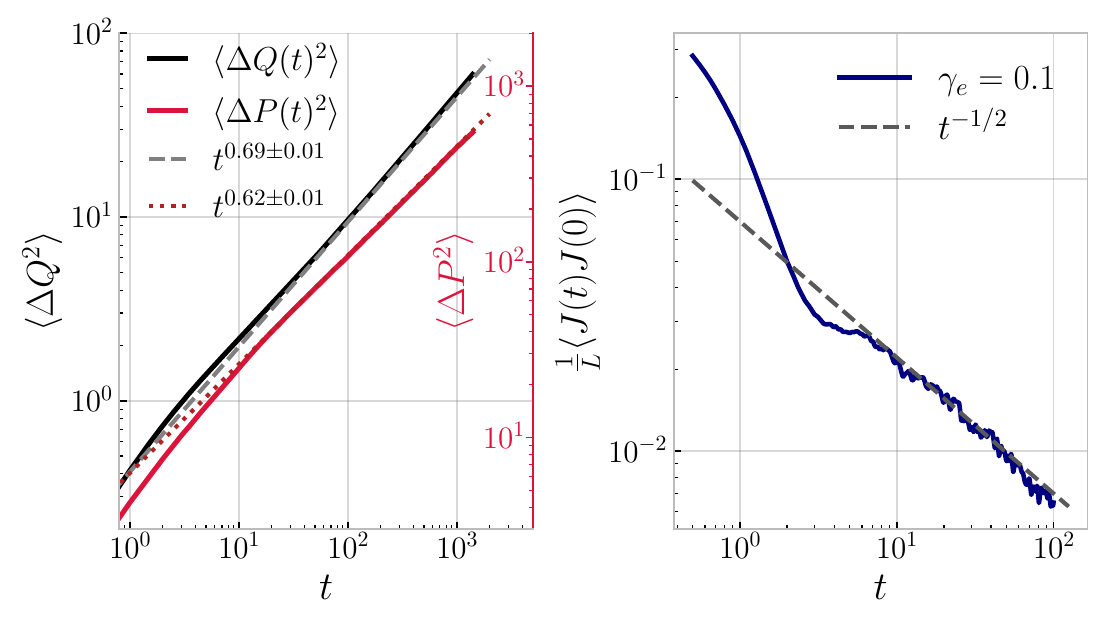}
    \caption{\textbf{Pure number/energy breaking} ($\gamma_p=0$).
(Left) Variances of equilibrium charge (black) and momentum transfer (red), $\langle \Delta Q(t)^2\rangle$ and $\langle \Delta P(t)^2\rangle$, in the number-breaking, momentum-conserving regime (here $\gamma_e=0.1$). Over the accessible time window the data show effective power-law growth with exponents $\alpha\approx 0.69$ for charge transfer and $\alpha\approx 0.62$ for momentum transfer (guide lines), consistent with an intermediate-time regime between the linear hydrodynamics estimate ($\Delta Q \sim t^{3/8}$, $\Delta P \sim t^{1/4}$), and the expected KPZ scaling ($\Delta Q\sim \Delta P \sim t^{1/3}$).
(Right) Total-current autocorrelator $L^{-1}\langle J(t)J(0)\rangle$, decaying approximately as $t^{-1/2}$, indicating nonintegrable long-time tails and superdiffusive charge transport in this regime. These simulations used $L=1000$.}
    \label{fig:num_breaking_only}
\end{figure}

\sectitle{Numerical evidence}---We introduce a stochastic multicomponent one-dimensional ideal gas that realizes the same linear fluctuating hydrodynamics as symmetry-broken graphene at charge neutrality, Eq.~\eqref{basichydro}. Particle number plays the role of energy density $e$ in Eq.~\eqref{basichydro}, while energy conservation in the stochastic gas is strongly broken~\footnote{A generic fluid conserving particle number, momentum, and energy also has two sound modes and a zero-velocity heat mode $h$. This is nevertheless a different nonlinear hydrodynamic problem: the heat mode generically couples to the sound modes $\phi_\pm$ as $j_h \sim \phi_+^2-\phi_-^2$ and is therefore already superdiffusive in the fully conserved theory, with the familiar $z=5/3$ scaling~\cite{SpohnNLFH,Delacretaz2020}. By contrast, such a coupling is forbidden for the charge mode at neutrality.}. Charge is represented by an additional label $q_i \in \{-1,0,+1\} $ carried by the particles. Particles move ballistically between collisions, where the relative velocity is stochastically redrawn while preserving ideal-gas static correlations. Particle number is relaxed by momentum-conserving particle splitting and merging events at rate $\gamma_e$, while momentum is relaxed by stochastic center-of-mass velocity flips at rate $\gamma_p$. Charge is conserved exactly, while charge labels are randomly mixed at collisions with probability $p_{\rm sw}$, generating a tunable Fickian contribution. These dynamics preserve a grand-canonical ideal-gas stationary state, which we take at density $\rho=1$ and temperature $k_BT=1$. The hydrodynamic rates $\gamma_{e,p}$ differ from the microscopic probabilities $(p_{\rm merge},p_{\rm flp})$ by nonuniversal $O(1)$ factors determined empirically; details are given in the Supplemental Material~\cite{suppmat}.

We first consider weakly breaking both number and momentum conservation. Holding the ratio $\gamma_e/\gamma_p$ fixed and taking the weak-dissipation limit, we observe that the charge diffusion constant exhibits a singular dependence on the symmetry-breaking rates. Two representative ratios are shown in Fig.~\ref{fig1}(b), illustrating that the diffusion constant can be either enhanced or suppressed relative to the fully conserved case. Correspondingly, the current-current correlator has a positive (correlated) or negative (anti-correlated) long-time tail (Fig.~\ref{fig1}(c)). We further scan over different ratios $\gamma_e/\gamma_p$ at weak symmetry breaking and extract the charge diffusion constant. The numerical results are consistent with the predicted singular square-root dependence in Eq.~\eqref{mainresult} (Fig.~\ref{fig1}(d)).

Finally, we consider pure energy relaxation ($\gamma_p=0$). In this regime momentum is conserved and eventually enters the strong-coupling regime described by Eq.~\eqref{coupledburgers}. Numerically, we observe superdiffusive charge transfer with variance growing as $t^{\alpha_Q}$ with $\alpha_Q \approx 0.69$, while the momentum transfer variance grows with $\alpha_P \approx 0.62$, consistent with recent numerical results for degenerate coupled-Burgers equations~\cite{Schmidt_2025}. Both charge and momentum transfer variances are shown in Fig.~\ref{fig:num_breaking_only} (left).

This behavior can be understood as an intermediate-time regime. At early times, the momentum mode is described by the linearized hydrodynamics and appears diffusive. The convective contribution to the charge transfer then scales superdiffusively with $\Delta Q(t)^2 \sim X(t)\sim t^{3/4}$ (from repeated returns of backscattered sound waves). At longer times, however, momentum and charge are governed by the coupled Burgers equations in Eq.~\eqref{coupledburgers}. In this regime, both momentum and charge transfer variances are expected to scale as $t^{2/3}$. One therefore expects intermediate-time effective exponents between $1/2$ and $2/3$ for momentum transfer and between $3/4$ and $2/3$ for charge transfer~\footnote{An alternative asymptotic scenario was recently proposed in Ref.~\cite{Schmidt_2025}, in which the two modes retain unequal dynamical exponents $z_1<3/2$ and $z_2>3/2$, rather than both approaching $z=3/2$. The present finite-time numerics do not distinguish between these possibilities.}, consistent with the exponents observed in Fig.~\ref{fig:num_breaking_only} (left). Figure~\ref{fig:num_breaking_only} (right) shows the corresponding current--current correlator, which decays approximately as $t^{-1/2}$, consistent with nonintegrable long-time tails and superdiffusive charge transport.


\sectitle{Discussion}---We have presented a hydrodynamic argument and unambiguous numerical evidence suggesting that the charge diffusion constant in a particle-hole symmetric fluid changes discontinuously upon adding perturbations that relax energy and/or momentum. Moreover, the diffusion constant in the weak noise limit takes a nonuniversal value that depends on the \emph{ratio} of energy and momentum relaxing processes. An important consequence of this is that techniques using zero-noise extrapolation on the d.c.~diffusivity, applied to this setting, will yield answers that are unrelated to the true zero-noise value. Thus our result illustrates a potential failure mode of zero-noise extrapolation techniques that are routinely used in classical and quantum simulations. Of course, in the present case, the solution is straightforward---one has to compute the time-dependent diffusion constant and extrapolate it suitably---but requires detailed knowledge of the underlying hydrodynamic theory. 

Here we have primarily focused on systems with PHS, since this simplifies the hydrodynamics in several ways. In particular, PHS is a sufficient condition for the mode structure required for hydrodynamic recoupling, since certain mode-coupling constants must vanish. More generally, however, a discrete symmetry is not required either for convective diffusion~\cite{Krajnik2024} or for the singular charge transport studied here. For locally equilibrated multicomponent fluids with a weakly relaxed sound sector, charge neutrality alone is sufficient for hydrodynamic recoupling. We give an explicit construction without assuming any discrete symmetry in the End Matter.

A natural question is whether our results can be extended to two dimensions, where, again, a random walker returns infinitely often to the origin. In this setting, the analysis is complicated by shear modes, which give rise to singularities in the diffusion constant even absent momentum or energy relaxation~\cite{Levchenko2026}. Whether signatures of convective and recoupling mechanisms can be found in higher dimensions (e.g., in nonlinear response~\cite{Kirkinis2026}) is an interesting question for future work.

It would be interesting to test our predictions experimentally in a range of settings, including multicomponent cold-atom gases~\cite{PhysRevLett.88.160401,Patel2023,Yan2024}, colloidal fluids, compensated electron-hole fluids and semimetals, graphene ribbons, and systems with long-lived spin, valley, or other flavor imbalances.
An especially interesting open question (which is beyond the scope of our methods) is how the convective effects we discuss change in the low-temperature limit (for example in neutral quantum-critical fluids) where the dominant energy fluctuations are quantum rather than thermal in character.

Finally, recent work~\cite{2026arXiv260722527K} develops a field-theory treatment of coupled nonlinear hydrodynamics. While this framework does not directly address the weak symmetry breaking considered here, it would be interesting to explore hydrodynamic recoupling from this perspective.

\sectitle{Acknowledgments}---The authors thank Jacopo De Nardis, Enej Ilievski, and Luka Paljk for helpful discussions. S.G. acknowledges support from NSF DMR-2516303. R.V.~acknowledges support from the Swiss National Science Foundation (grant 10008234) and the Foundation for the University of Geneva (FUNIGE). E.M. acknowledges support from the European Union (ERC, STAQQ, 101171399). Views and opinions expressed are however those of the authors only and do not necessarily reflect those of the European Union or the European Research Council Executive Agency. Neither the European Union nor the granting authority can be held responsible for them. The simulations presented in this article were performed on computational resources managed and supported by Princeton Research Computing, a consortium of groups including the Princeton Institute for Computational Science and Engineering (PICSciE) and Research Computing at Princeton University.

\makeatletter
\let\origbibsection\bibsection
\renewcommand{\bibsection}{%
  \begingroup
    \let\addcontentsline\@gobblethree
    \origbibsection
  \endgroup
}
\makeatother

\bibliography{Refs}

\clearpage

\twocolumngrid
\begin{center}
    \textbf{\large End Matter}
\end{center}

\sectitle{Charge diffusion constant}---The third line of Eq.~\eqref{basichydro} expresses the constitutive relation for the charge current. By particle-hole symmetry the two leading symmetry-allowed terms are the Fick's law term (with coefficient $D_{\rm Fick}$, and accompanying noise $\eta$) and a nonlinear ``convective'' coupling of charge to sound waves. Since (at this order) the motion of sound waves is unaffected by charge fluctuations, these fluctuations can be regarded as a source of extrinsic multiplicative noise. 
The charge diffusion constant can be defined in terms of the autocorrelation function of the charge density: $2\mathfrak{D}\,t \simeq \chi^{-1}\sum_x x^2\,\langle n(x,t)\,n(0,0)\rangle$ (where $\chi$ is the charge susceptibility). Here, the stochastic average $\langle \ldots \rangle$ involves two sources of noise: (i)~the noise term $\eta$ and (ii)~the multiplicative ``velocity'' noise that arises because sound waves advect thermal fluctuations of energy and momentum density, imparting random kicks to the charge pattern.

We will now compute this diffusion constant in two steps. First, we will artificially set $D_{\rm Fick}=0$. In this limit, the density obeys the first-order differential equation $\partial_t n+\partial_x(pn)=0$. This equation can be solved by dividing the density into small packets, each with an evolving position: $n(x,t)=\int dx_0\, n(x_0,0)\,\delta(x-X(x_0,t))$, with characteristics solving $\partial_t X=p(X(x_0,t),t)$. A crucial simplification is that the ballistic sound waves move parametrically faster than the packets of charge; therefore, one can replace $p(X,t)$ with $p(x_0,t)$ up to subleading corrections~\cite{pnas.2403327121}. Making this approximation, one arrives at a rigid shift of the charge pattern, 
\beq\label{eq:rigid_shift}
n(x,t) \simeq n_0(x-X(t)), \quad X(t)\equiv \int_0^t d\tau\,p(x_0,\tau),
\eeq
where $p(x_0,\tau)$ is obtained by propagating the momentum along characteristics $p(x,t)=p_r(x-ct)+p_l(x+ct)$. The structure factor is then given by
\beq
\langle n(x,t) n(0,0)\rangle
\simeq\hspace{-0.25mm}
\int\hspace{-0.5mm} dX P_t(X)\langle n_0(x-X) n_0(0)\rangle
\simeq
\chi P_t(x),
\label{variance}
\eeq
where $P_t(x)\equiv\langle\delta(x-X(t))\rangle$ is the distribution of the random displacement $X(t)$. Therefore, in the pure convection limit, the charge diffusion constant $\mathfrak{D}=D_{\rm conv}$ is equal to the diffusion constant of random displacement $X$ (via the second moment of the structure factor).


To complete the calculation, we restore Fickian processes, as follows. Consider conditioning on a particular history of sound waves and a particular initial charge configuration. On this history, the charge pattern is rigidly translated by $X(t)$. Beyond this rigid translation, the charge dynamics is due to Fickian processes. If one treats the elements of charge as particles, one can describe these two effects as follows: 
the particle configuration is randomly rearranged over the scale $\sqrt{D_{\rm Fick} t}$ in addition to being shifted by $\sqrt{D_{\rm conv} t}$.
The full diffusion constant comes from convolving these effects~\cite{pnas.2403327121,yoshimura2025hydrodynamicfluctuationsstochasticcharged}, $n(x,t) = \int dx' G_{\rm Fick}(x-x',t) n_0(x'-X(t))$: 
for each fixed $X$, the momentum-space charge structure factor at momentum $k$ is $S_{X}(k,t) = \chi \exp[-D_{\mathrm{Fick}} k^2 t + i k X(t)]$; averaging this structure factor over the Gaussian variable $X(t)$ gives the final result $S(k,t) = \chi \exp[-(D_{\mathrm{Fick}} + D_{\mathrm{conv}}) k^2 t]$, which yields Eq.~\eqref{eq1}. 
Since the convective and Fickian effects can be treated separately, we will focus below on the fate of the convective contribution when relaxation processes are allowed; the Fickian contribution can be restored at the end of the calculation.

\sectitle{Momentum-breaking limit}---In this section, we consider setting $\gamma_e = 0$, for example by considering static disorder instead of noise. In this case, energy evolves diffusively, with a diffusion constant set by $D_e \sim 1/\gamma_p$. 
By Lorentz covariance, momentum density is equal to the energy current: consequently, $X(t)$ is the integrated current of a diffusive quantity across a fixed point, and as such scales as $X(t) \sim t^{1/4}$. 
Thus, the convective contribution to charge transport becomes \emph{subdiffusive} with dynamical exponent $z=4$ for any nonzero $\gamma_p$ when $\gamma_e = 0$. In this regime, the convective contribution to charge transport is subleading to the Fickian term, and the diffusion constant is simply $D_{\rm Fick}$ as $\gamma_p \to 0$. 
Intuitively, the subdiffusive behavior can be understood as follows (see also~\cite{PhysRevLett.127.230602,10.1073/pnas.2202823119}). In time $t$, a given packet of charge is repeatedly encountered by the same sound waves as they scatter back and forth. Since the backscattering of sound waves conserved energy, its momentum changes sign at each backscatter, so that successive encounters with a wave give opposite kicks. As a result, each sound wave contributes only an $O(1)$ random kick to the net displacement of a charge packet over time $t$. Since the sound waves themselves diffuse, only $O(t^{1/2})$ independent sound waves cross the origin in time $t$. As a sum of $t^{1/2}$ random kicks, the net displacement grows as $X(t)\sim t^{1/4}$.
This limit clearly illustrates the crucial difference between the $\gamma_p = 0$ and $\gamma_p \neq 0$ cases: in the former case each charge packet encounters each sound wave exactly once (by hydrodynamic decoupling), but for any nonzero $\gamma_p$ the charge packet encounters each sound wave \emph{infinitely many times} (in quasi-1D geometries).

\begin{figure}    \includegraphics[width=0.98\columnwidth, trim=0mm 4mm -0.5mm 1mm,
  clip]{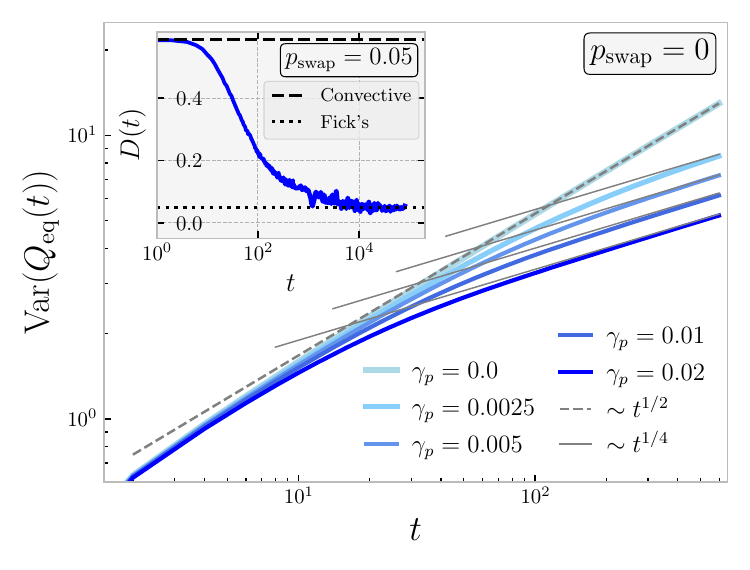}
    \caption{\textbf{Momentum breaking} ($\gamma_e=0$).
(Main panel) Equilibrium charge-transfer variance ${\rm Var}(Q_{\rm eq}(t))$ with no charge-label swaps ($p_{\rm sw}=D_{\rm Fick}=0$). For $\gamma_p=0$ the transfer is diffusive at long times, while any $\gamma_p>0$ produces a crossover on the timescale $\gamma_p^{-1}$ to subdiffusive growth ${\rm Var}(Q_{\rm eq})\sim t^{1/4}$ (guide lines).
These simulations used system length $L=1500$.
(Inset) Restoring charge-label swaps ($p_{\rm sw}=0.05$) yields a time-dependent diffusion constant $D(t)$ with two plateaus: an early-time plateau corresponding to both the Fickian and convective contributions, and a late-time plateau set by the bare Fickian diffusion constant. This simulation used $L=6000$.}
    \label{fig:mom_breaking_only}
\end{figure}

We check our predictions in this limit $\gamma_e=0$ in Fig.~\ref{fig:mom_breaking_only}. To isolate the convective effect, we first eliminate Fickian charge diffusion by disallowing any charge-label swaps. In this limit, $\gamma_p$ controls a crossover from diffusive to subdiffusive charge transfer, as shown in Fig.~\ref{fig:mom_breaking_only} (main panel). At short times the charge transfer grows diffusively, because sound packets propagate ballistically for long times before backscattering and therefore impart effectively uncorrelated kicks. The crossover to subdiffusive behavior occurs on the timescale $\gamma_p^{-1}$, when a sound packet revisits the same charge parcel with reversed velocity, producing kicks of opposite sign. We then restore Fickian diffusion by allowing charge-label exchanges. The resulting time-dependent diffusion constant exhibits two plateaus: an initial one corresponding to the fully conserved dynamics and a late-time plateau set by the bare Fickian contribution (Fig.~\ref{fig:mom_breaking_only} inset).

\sectitle{Charge transfer statistics}---While our discussion of charge transport in the main text has centered on the charge structure factor, we note that equilibrium charge transfer across a cut provides an equivalent probe of the charge diffusion constant~\cite{PhysRevLett.131.210402,wienand2023emergence}: the asymptotic growth of the charge-transfer variance obeys~\cite{suppmat} $\langle \Delta Q(t)^2\rangle \simeq 2\chi\sqrt{D t/\pi}$. We briefly comment on how higher cumulants of charge transfer cross over as energy and momentum relaxation is introduced. For standard diffusion, these higher cumulants scale with the usual macroscopic fluctuation form $t^{1/2}$~\cite{DerridaConjecture,Roche_2005,Bertini_2015, PhysRevLett.131.210402}, while for particle--hole symmetric diffusion the $n$-th cumulant scales as $t^{n/4}$~\cite{Krajnik2022a,Krajnik2022b,Krajnik2022c,pnas.2403327121,PhysRevE.111.015410,PhysRevE.111.024141,yoshimura2025hydrodynamicfluctuationsstochasticcharged}. This enhancement of charge-transfer fluctuations reflects the spatiotemporal correlations in the effective noise ``bath'' generated by sound waves. When energy and momentum relaxation are both present, one generically expects a crossover to the standard $t^{1/2}$ scaling, on a timescale set by the slower of the two relaxation rates. When only energy is conserved and $D_{\rm Fick}=0$, the higher moments remain anomalous at all times, as computed in Refs.~\cite{pnas.2403327121,PhysRevE.111.015410,PhysRevE.111.024141,yoshimura2025hydrodynamicfluctuationsstochasticcharged}. The dynamics of higher moments in the complementary limit where only momentum is conserved is more subtle, and we defer a detailed analysis to future work.


\sectitle{Beyond particle-hole symmetry}---We now show that hydrodynamic recoupling does not require PHS. Consider a two-component gas with separately conserved species densities $\rho_\pm$, and define the
total density and species imbalance
\begin{equation}
    \rho=\rho_++\rho_-,
    \qquad
    n=\rho_+-\rho_-.
\end{equation}
We assume no symmetry exchanging the two species, and consider a
charge-neutral reference state with $n_0=0$. In equilibrium, the Euler currents of the species densities
take the form $j_\pm=\rho_\pm u$, where $u$ is the local fluid velocity~\cite{deBoer2018PerfectFluids,deBoer2020NonBoost}. This result does not require boost
symmetry; a simple derivation from the local Gibbs ensemble is given in
the Supplemental Material~\cite{suppmat}. It follows immediately that
\begin{equation}
    j_\rho=\rho u,
    \qquad
    j_n=n u.
\end{equation}
Thus, at charge neutrality the Euler-scale charge current vanishes
identically when $n=0$, to all orders in fluctuations of the other
hydrodynamic fields. Expanding about a zero-flow state,
$u=p/\chi_p+\ldots$, its leading contribution is therefore $j_n=\chi_p^{-1}np+\ldots$,
precisely the convective coupling to momentum discussed in the main text.

To match the three-mode hydrodynamics of the main text, we take energy to be rapidly relaxed by a local bath at temperature $T$ (which
preserves the species densities and momentum). Energy is therefore not a
hydrodynamic mode, while the Euler currents above are unchanged; see the
Supplemental Material~\cite{suppmat}.

Without PHS, the mode that couples to momentum to form sound need not coincide with the total-density fluctuation. In general, at fixed temperature and to linear order, the pressure varies as $\delta P=P_\rho\delta \rho+P_n n$. 

Defining 
\begin{equation}
    \hat{\rho}=\delta \rho+(P_n/P_\rho) n,
\end{equation}
so that $\delta P = P_\rho \hat \rho$, the linear Euler equations become
\begin{equation}
    \partial_t\hat \rho+\rho_0\chi_p^{-1}\partial_xp=0,
    \quad
    \partial_t p+P_\rho\partial_x\hat \rho=0,
    \quad
    \partial_t n=0.
\end{equation}
Thus $\hat\rho$ and $p$ form the two sound modes, while $n$ remains the
diffusive charge mode. PHS would simply enforce $P_n=0$, so that
$\hat \rho=\delta \rho$ and this mode decomposition follows directly from
symmetry.

Finally, weakly relax total particle number and momentum while preserving species imbalance
$n$. For reservoirs matched to the reference equilibrium state, the local
Gibbs ensemble obeys the Gibbs--Duhem relation
\begin{equation}
    dP=s\,dT+\rho\,d\mu_\rho+n\,d\mu_n+p\,du .
\end{equation}
At fixed temperature, and expanding about the neutral zero-flow state
$n_0=p_0=0$ gives $\delta P=\rho_0\,\delta\mu_\rho=P_\rho \hat\rho$. Thus a particle-number bath that relaxes the chemical-potential mismatch
$\delta\mu_\rho$ damps precisely the pressure-carrying field $\hat\rho$, while
$n$ remains conserved.

The resulting long-wavelength equations are
\begin{align}
    \partial_t\hat \rho+\rho_0\chi_p^{-1}\partial_x p+\ldots
        &= -\gamma_f\rho\hat \rho, \\
    \partial_t p+P_\rho\partial_x\hat \rho+\ldots
        &= -\gamma_p p, \\
    \partial_t n+\chi_p^{-1}\partial_x(np)
        &=D_n\partial_x^2 n+\ldots .
\end{align}
These equations have the same form as those studied in the main text. The singular weak-relaxation dependence of the charge diffusion constant, therefore carries over unchanged. Further details and the microscopic derivation of the Euler currents are given in the Supplemental Material~\cite{suppmat}.

\newpage

\widetext

\makeatletter
\begin{center}
\textbf{\large Supplementary Materials: Anomalous diffusion and singular transport from hydrodynamic recoupling}

\vspace{3mm}

Ewan McCulloch,\textsuperscript{1,2} Romain Vasseur,\textsuperscript{3} Sarang Gopalakrishnan\textsuperscript{2}

\vspace{2mm}

\textsuperscript{1}\textit{\small Laboratoire de Physique de l'École Normale Supérieure, CNRS,\\
ENS \& Université PSL; 24 rue Lhomond, 75005 Paris, France}

\textsuperscript{2}\textit{\small Department of Electrical and Computer Engineering,\\ Princeton University, Princeton, NJ 08544, USA}

\textsuperscript{3}\textit{\small Department of Theoretical Physics, University of Geneva,\\ 24 quai Ernest-Ansermet, 1211 Gen\`eve, Switzerland}

\makeatother


\makeatother

\end{center}

\tableofcontents

\setcounter{equation}{0}
\setcounter{figure}{0}
\setcounter{table}{0}
\setcounter{page}{1}
\makeatletter
\renewcommand{\theequation}{S\arabic{equation}}
\renewcommand{\thefigure}{S\arabic{figure}}

\section{Charge transfer variance and the diffusion constant}

In this appendix we relate the long-time growth of equilibrium charge-transfer fluctuations across a cut to the equilibrium diffusion constant $D$. We consider a conserved charge density $n(x,t)$ in equilibrium, obeying the continuity equation $\partial_t n(x,t) + \partial_x j(x,t) = 0$, with static susceptibility $\chi \equiv \int dx\, \langle n(x,0)n(0,0)\rangle$. We assume that the charge density is transported diffusively with diffusion constant $D$. That is to say, we assume that the retarded density--density correlator has a
diffusive pole,
\begin{equation}
G^R_{nn}(k,\omega)
\xrightarrow[k,\omega\to 0]{}
\frac{\chi D k^2}{-i\omega + Dk^2}.
\label{eq:diffusive_pole}
\end{equation}
This assumption specifies the universality class of transport but does not impose a microscopic constitutive relation such as Fick’s law. In particular, it allows for non-Gaussian full counting statistics arising from nonlinear or convective couplings to additional slow modes, since such effects enter only through higher cumulants.

Using charge conservation, $n(k,\omega)=\frac{k}{\omega}j(k,\omega)$, the diffusive pole \eqref{eq:diffusive_pole} implies $\mathrm{Im}\, G^R_{jj}(k,\omega) = \chi D\,\omega + O(\omega k^2)$ as $k,\omega\to 0$. It follows immediately that~\cite{Kubo,forster2018hydrodynamic}
\begin{equation}
D = \frac{1}{\chi}
\lim_{\omega\to 0}\lim_{k\to 0}
\frac{1}{\omega}\,
\mathrm{Im}\, G^R_{jj}(k,\omega),
\end{equation}
so that the diffusion constant entering the pole of
$G^R_{nn}$ is precisely the diffusion constant defined by the Kubo formula~\cite{Kubo}.

We define the integrated charge transfer across the origin,
$\Delta Q(t) \equiv \int_0^t d\tau\, j(0,\tau)$. Using the continuity equation, this may be written exactly as $\Delta Q(t)
= \int_0^\infty dx\,\big[n(x,t)-n(x,0)\big]$. The variance of the equilibrium charge transfer is therefore
\begin{equation}
\langle \Delta Q(t)^2\rangle
= 2 \int_0^\infty dx \int_0^\infty dx'\,
\Big[
\langle n(x,0)n(x',0)\rangle
- \langle n(x,t)n(x',0)\rangle
\Big].
\end{equation}

Using Eq.~\eqref{eq:diffusive_pole}, the long-wavelength density
correlator takes the form $\langle n(x,t)n(0,0)\rangle
= \chi \int\frac{dk}{2\pi}\, e^{ikx-Dk^2 t}$.
Substituting this into the expression for the charge-transfer variance and performing the spatial integrals gives
\begin{equation}
\langle \Delta Q(t)^2\rangle
= 2\chi \int_0^\infty \frac{dk}{\pi}\,
\frac{1-e^{-Dk^2 t}}{k^2}.
\end{equation}
Evaluating the remaining integral yields the universal asymptotic form~\cite{DerridaConjecture,Roche_2005,Bertini_2015, PhysRevLett.131.210402}
\begin{equation}
\langle \Delta Q(t)^2\rangle
\underset{t\to\infty}{\sim} \frac{2\chi}{\sqrt{\pi}}\,\sqrt{D t}.
\end{equation}

\section{Singular charge diffusion from linear hydrodynamics with symmetry breaking}
\label{app:linear_hydro}

In the main text we introduced an intuitive picture of charge transport based on isolated ballistically propagating sound waves that are stochastically backscattered by weak symmetry-breaking processes. The purpose of this appendix is to provide a complementary derivation of the same result within linear fluctuating hydrodynamics. While the toy model emphasizes individual ballistic excursions, the hydrodynamic calculation encodes their cumulative effect through long-wavelength equilibrium correlations.

We focus on the convective contribution to equilibrium charge transport, which arises from the nonlinear coupling between charge and the energy current. At charge neutrality this provides a distinct contribution to equilibrium charge-transfer fluctuations, in addition to ordinary Fickian diffusion. In the weak symmetry-breaking regime, this contribution may be isolated by a decoupling assumption as described in the main text: to leading order in the symmetry-breaking rates, the sound modes evolve independently and act to advect the entire charge profile by a stochastic displacement~\cite{pnas.2403327121}.

Under this assumption, the convective contribution can be obtained by probing the momentum (energy current) $p$ at a fixed spatial location. The resulting stochastic displacement is
\begin{equation}
X(t) = \int_0^t dt'\, p(0,t').
\end{equation}
The long-time statistics of $X(t)$ determine the convective contribution to the equilibrium charge-transfer variance and hence to the equilibrium charge diffusion constant.

\subsection{Linear fluctuating hydrodynamics with symmetry breaking}

We consider linear fluctuating hydrodynamics for graphene energy transport in one dimension, formulated in terms of the energy density $e(x,t)$ and the energy current $p(x,t)$. (The stochastic gas model introduced earlier provides a minimal microscopic realization of the same hydrodynamic structure, with $e$ and $p$ identified respectively with particle number and momentum density.) Weak symmetry breaking relaxes both energy and momentum, while diffusive corrections encode gradient-scale dissipation. The equations are
\begin{align}
\partial_t e(x,t) + \partial_x p(x,t)
&= -\gamma_e\, e(x,t) + \eta_e(x,t),
\label{eq:eps_hydro_app} \\
\partial_t p(x,t) + c^2 \partial_x e(x,t)
&= -\gamma_p\, p(x,t)
+ D_p \partial_x^2 p(x,t)
+ \eta_p^{\rm OU}(x,t)
+ \partial_x \eta_p^{\rm cons}(x,t),
\label{eq:phi_hydro_app}
\end{align}
where $c$ is the sound velocity, $\gamma_e$ and $\gamma_p$ are the symmetry-breaking relaxation rates, and $D_p$ is the Fickian diffusion constant for the energy current. The noise correlators are fixed by the fluctuation-dissipation theorem~\cite{landau1987fluid}, and are given by
\begin{align}
\langle \eta_e(x,t)\,\eta_e(x',t') \rangle
&= 2 \gamma_e \chi_e\,
\delta(x-x')\delta(t-t'), \nonumber \\
\langle \eta_p^{\rm OU}(x,t)\,\eta_p^{\rm OU}(x',t') \rangle
&= 2 \gamma_p \chi_p\,
\delta(x-x')\delta(t-t'),\nonumber \\
\langle \eta_p^{\rm cons}(x,t)\,\eta_p^{\rm cons}(x',t') \rangle
&= 2 D_p \chi_p\,
\delta(x-x')\delta(t-t'),\label{eqn:noise}
\end{align}
with all cross-correlators vanishing. These choices ensure a Gaussian stationary state. Eqs.~\eqref{eq:eps_hydro_app}–\eqref{eq:phi_hydro_app} constitute the most general linear fluctuating hydrodynamic theory for energy density and energy current consistent with translation invariance, time-reversal symmetry, and spatial parity. (At linear order, no additional gradient terms are allowed in the energy continuity equation since $p$ is defined as the energy current.)

Fourier transforming Eqs.~\eqref{eq:eps_hydro_app}–\eqref{eq:phi_hydro_app} in space and time yields
\begin{equation}
\begin{pmatrix}
-i\omega + \gamma_e & ik \\
ic^2 k & -i\omega + \gamma_p + D_p k^2
\end{pmatrix}
\begin{pmatrix}
e(k,\omega) \\ p(k,\omega)
\end{pmatrix}
=
\begin{pmatrix}
\eta_e(k,\omega) \\
\eta_p^{\rm OU}(k,\omega) + ik\,\eta_p^{\rm cons}(k,\omega)
\end{pmatrix}.
\label{eq:hydro_matrix}
\end{equation}

Inverting this matrix, we obtain
\begin{equation}
p(k,\omega)
=
\frac{
ic^2 k\, \eta_e
+ (-i\omega + \gamma_e)
\bigl(\eta_p^{\rm OU} + ik\,\eta_p^{\rm cons}\bigr)
}{
(-i\omega + \gamma_e)(-i\omega + \gamma_p + D_p k^2)
+ c^2 k^2
}.
\end{equation}

Using the noise correlators above, the energy-current structure factor takes the form
\begin{equation}
S_{pp}(\omega,k)\equiv \langle p(k,\omega)p(-k,-\omega)\rangle
=
\frac{
2\gamma_e c^4 \chi_e k^2
+
2(\gamma_p + D_p k^2)\chi_p
\bigl(\omega^2 + \gamma_e^2\bigr)
}{
\left[
\omega^2
- c^2 k^2
- \gamma_e(\gamma_p + D_p k^2)
\right]^2
+
\omega^2(\gamma_e + \gamma_p + D_p k^2)^2
}.
\label{eq:Sphiphi}
\end{equation}

As discussed in the main text and above, the long-time growth of $\langle X(t)^2\rangle$ defines the convective contribution to the equilibrium charge diffusion constant. Using translational invariance,
\begin{equation}
\langle X(t)^2\rangle
=
\int_0^t dt_1 \int_0^t dt_2
\int \frac{dk\, d\omega}{(2\pi)^2}
\, e^{-i\omega(t_1-t_2)}
S_{pp}(\omega,k).
\end{equation}
In the long-time limit, the double time integral yields
\beq\label{nascent}
\int_0^t dt_1 \int_0^t dt_2\, e^{-i\omega(t_1-t_2)}
\;\xrightarrow[t\to\infty]{}\;
2\pi t\, \delta(\omega),
\eeq
so that $\langle X(t)^2\rangle \simeq 2 D t$, with $D$ given by 
\begin{equation}\label{eqn:DX}
D(\gamma_e,\gamma_p)
=
\int \frac{dk}{4\pi}\,
S_{pp}(k,0) =
\int \frac{dk}{4\pi}\,\frac{
2\gamma_e
\bigl(
c^4 \chi_e k^2
+
\chi_p \gamma_e(\gamma_p + D_p k^2)
\bigr)
}{
\bigl[
\gamma_e(\gamma_p + D_p k^2)
+ c^2 k^2
\bigr]^2
}.
\end{equation}
Performing this integral exactly gives,
\beq\label{eq:DX_exact}
D(\gamma_e,\gamma_p)
=\frac{1}{4}\,\sqrt{\frac{\gamma_e}{\gamma_p}}\;
\frac{c^4\chi_e+\chi_p\bigl(c^2+2\gamma_e D_p\bigr)}
{\bigl(c^2+\gamma_e D_p\bigr)^{3/2}}.
\eeq

\subsection{Double scaling limit}
We now evaluate Eq.~\eqref{eq:DX_exact} in the joint weak-dissipation limit where both symmetry-breaking rates are taken to zero at fixed ratio, $\gamma_e=r\,\gamma_p$ with $r>0$ fixed, and $\gamma_e\to0$. In this limit $\gamma_e D_p \ll c^2$, and the exact expression simplifies,
\begin{equation}
D(\gamma_e,\gamma_p)
=
\frac{c^2\chi_e+\chi_p}{4c}\sqrt r
\left(1+O\!\Bigl(\frac{\gamma_e D_p}{c^2}\Bigr)\right).
\end{equation}
Thus, to leading order in the double scaling limit,
\begin{equation}
D(\gamma_e,\gamma_p)
\;\xrightarrow[\gamma_e,\gamma_p\to0]\;
\frac{c^2\chi_e+\chi_p}{4c}\sqrt{\frac{\gamma_e}{\gamma_p}}.
\label{eq:DX_final}
\end{equation}
Equation~\eqref{eq:DX_final} shows that in the weak-dissipation limit the convective contribution approaches a finite value that depends on the ray
$r=\gamma_e/\gamma_p$ along which the fully conserved point is approached, reflecting
the noncommutativity of the limits $\gamma_e\to0$ and $\gamma_p\to0$.

\subsection{Fully conserved case}
We now evaluate the convective contribution in the fully conserved limit
$\gamma_e=\gamma_p=0$, setting $\gamma_e=\gamma_p=0$ directly in  Eq.~\eqref{eq:Sphiphi}. The variance of the walk $X(t)$ is then given by
\begin{equation}
\langle X(t)^2\rangle
=\int_0^t dt_1\int_0^t dt_2
\int\frac{dk\,d\omega}{(2\pi)^2}\,
e^{-i\omega(t_1-t_2)}\,\frac{2D_p\,\chi_p\,k^2\,\omega^2}
{\left(\omega^2-c^2k^2\right)^2+\omega^2\left(D_p k^2\right)^2}.
\label{X2_start}
\end{equation}
Performing the $k$ integral gives
\begin{equation}
I(\omega)\equiv \int_{-\infty}^{\infty}\frac{dk}{2\pi}\,S_{pp}(\omega,k)=\frac{\chi_p}{c}\,
\left(\frac{1+\sqrt{1+\left(\frac{D_p|\omega|}{c^2}\right)^2}}
{2\left[1+\left(\frac{D_p|\omega|}{c^2}\right)^2\right]}\right)^{1/2}.
\label{Iomega_def}
\end{equation}
Using the nascent delta function identity in Eq.~\eqref{nascent} gives $\langle X(t)^2\rangle \;\xrightarrow[t\to\infty]{} t\int\frac{d\omega}{2\pi}\, \delta(\omega)\,I(\omega)$. Since $I(0)=\chi_p/c$, we obtain $\langle X(t)^2\rangle \simeq \frac{\chi_p}{c}\,t$, and hence, comparing with $\langle X(t)^2\rangle\simeq 2D_{\rm conv}t$, the convective diffusion constant is
\beq
D_{\rm conv}=\frac{\chi_p}{2c}.
\label{Dconv_conserved}
\eeq
In the strictly conserved theory, the static susceptibilities $\chi_e$ and $\chi_p$ are not independent. Requiring a time-translation-invariant Gaussian equilibrium fixes the ratio of susceptibilities to the speed of sound. In our normalization this gives $\chi_p=c^2\chi_e$, so that Eq.~\eqref{Dconv_conserved} may be rewritten as $D_{\rm conv}=(c^2\chi_e+\chi_p)/4c$, i.e., the weak dissipation limit on the ray $\gamma_e=\gamma_p$.

Throughout this work we assume that the weak symmetry-breaking noise is implemented so as to respect the stationary ensemble of the fully conserved case (as in the deformed ideal gas studied in our numerics). Otherwise, the weak-dissipation limit and the fully conserved limit could differ for a trivial reason: the effective baths would impose different equal-time fluctuations (different $\chi_e,\chi_p$), and the diffusion constant would jump simply because the stationary state changes, rather than due to the non-Markovian effects that are the focus of this work.

\subsection{Charge structure factor in the energy conserving case}
Consider the rigid shift form of the charge density in Eq.~\eqref{eq:rigid_shift}. Its spatial Fourier transform gives
\begin{equation}
n_k(t)\equiv \int dx\,e^{-ikx}\,n(x,t)=e^{-ikX(t)}n_k(0).
\end{equation}
Assuming $X(0)=0$ (translational invariance), the corresponding dynamical structure factor is
\begin{equation}\label{eq:Snn_def}
S_{nn}(k,t)\equiv \langle n_k(t)n_{-k}(0)\rangle= \chi \big\langle e^{-ikX(t)}\big\rangle.
\end{equation}
For $\gamma_e=0$, $X(t)$ is sum of independent random kicks from ${\cal O}(\sqrt{t})$ sound waves, and is therefore a Gaussian random variable. The generating function in Eq.~\eqref{eq:Snn_def} takes the form
\begin{equation}\label{eq:Snn_gauss}
S_{nn}(k,t)=\chi \exp\!\left[-\frac{k^2}{2}\,\langle X(t)^2\rangle\right].
\end{equation}
For $\gamma_e=0$ and $D_p=0$ the long-time growth of $\langle X(t)^2\rangle$ is controlled by the slow diffusive
mode with diffusion constant $D_E=c^2/\gamma_p$, yielding
\begin{equation}\label{eq:X2_energy}
\langle X(t)^2\rangle
\simeq
\frac{2\chi_p}{c\sqrt{\pi}}\sqrt{t/\gamma_p}.
\end{equation}
Combining Eqs.~\eqref{eq:Snn_gauss} and \eqref{eq:X2_energy}, we obtain the stretched-exponential decay
\begin{equation}\label{eq:Snn_final}
S_{nn}(k,t)
\simeq \chi
\exp\!\left[
-\frac{\chi_p}{c\sqrt{\pi}}\sqrt{\frac{t}{\gamma_p}}\;k^2
\right].
\end{equation}
Equivalently, $n(x,t)$ is subdiffusive with dynamical exponent $z=4$. This structure factor is characteristic of the ``XNOR'' or ``tracer diffusion'' universality class discussed in Refs.~\cite{PhysRevLett.127.230602,10.1073/pnas.2202823119,PhysRevB.106.094303}, that we derived here from a purely hydrodynamic perspective.

\subsection{Optical conductivity}

Going back to the general case, the formalism outlined above can also be used to compute the optical conductivity $\sigma(\omega)$ given by the Kubo formula
\begin{equation}
\sigma(\omega) = \lim_{k \to 0}  \int_0^\infty dt {\rm e}^{i \omega t} S_{jj} (k,t),
\end{equation}
with $ S_{jj} (k,t)$ the current-current correlator. 
Using the continuity equation $\partial_t n + \partial_x j = 0$, we can compute the $k\to 0$ limit of the current-current correlator using the general form
\begin{equation}
S_{nn}(k,t)=\chi \exp\!\left[- D_{\rm Fick} k^2 t -\frac{k^2}{2}\,\langle X(t)^2\rangle\right].
\end{equation}
of the structure factor. Note that the variable $X(t)$ remains Gaussian in the general case, as it is a linear combination of Gaussian variables given by linearized hydrodynamics. We find
\begin{equation}
\sigma(\omega) = \chi D_{\rm Fick} + \frac{\chi}{2}   \int_0^\infty dt {\rm e}^{i \omega t} \frac{d^2}{dt^2} \langle X(t)^2\rangle ,
\end{equation}
where the first term corresponds to the Fickian contribution, whereas the second one is due to convective terms. 
Finally, using the general expression~\eqref{fourierform} and focusing on the real part of the conductivity, we obtain 
\begin{equation}
\sigma(\omega) = \chi D_{\rm Fick} + \chi \int \frac{dk}{4\pi} S_{pp}(k, \omega).
\end{equation}
In the d.c.~limit $\omega \to 0$, we recover $\sigma = \chi D$ with $D= D_{\rm Fick} + \int \frac{dk}{2\pi} S_{pp}(k, 0) $ the diffusion constant. 

\section{Exact quasiparticle representation of linear fluctuating hydrodynamics}

We show here that the toy model of reflected sound waves used in the main text is in fact an exact kinetic (Boltzmann) representation of the linear fluctuating hydrodynamics
Eqs.~\eqref{eq:eps_hydro_app}–\eqref{eq:phi_hydro_app}. In the absence of the viscous $D_p \partial_x^2 p$ term, the linearized equations reduce to a telegrapher-type dynamics for chiral sound packets~\cite{Goldstein1951,Kac1974ASM}. The sound packets can be represented in terms of quasiparticles which carry both a propagation direction and a particle/hole character, and the stochastic quasiparticle dynamics reproduces the hydrodynamic equations without approximation.

We introduce quasiparticle densities $f_{a,s}(x,t)$, where $s=\pm$ labels the propagation direction with velocity $v=s\,c$, and $a=\pm$ labels particle-like ($+$) versus hole-like ($-$) waves. The hydrodynamic fields are represented exactly as
\begin{equation}\label{eqn:qp-decomp}
e(x,t) = \sum_{a,s} a\, f_{a,s}(x,t),\quad 
p(x,t) = c \sum_{a,s} a\, s\, f_{a,s}(x,t).
\end{equation}

The kinetic evolution consists of ballistic propagation together with two independent symmetry-breaking Poisson processes, corresponding to the two relaxation rates in
Eqs.~\eqref{eq:eps_hydro_app}–\eqref{eq:phi_hydro_app}:
(i) a number/energy-breaking event, which converts particle/hole type and reverses direction, $(a,s)\xrightarrow{\gamma_e/2}(-a,-s)$, and (ii) a momentum-breaking event, which reverses direction without changing particle/hole type, $(a,s)\xrightarrow{\gamma_p/2}(a,-s)$. These processes are encoded by the linear kinetic equation
\begin{align}
\partial_t f_{a,s} + s c\,\partial_x f_{a,s}
&= -(\gamma_e+\gamma_p)/2\, f_{a,s}
+ \gamma_e/2\, f_{-a,-s}
+ \gamma_p/2\, f_{a,-s}
+ \zeta_{a,s}(x,t),
\label{eq:boltzmann_app}
\end{align}
where $\zeta_{a,s}$ is Gaussian white noise. Choosing the noise covariance so that the stationary ensemble for $e$ and $p$ is Gaussian with susceptibilities $\chi_e$ and $\chi_p$ yields exactly the fluctuation--dissipation relations in Eq.~\eqref{eqn:noise}, namely
\begin{equation*}
\langle \eta_a(x,t)\,\eta_a(x',t') \rangle
= 2 \gamma_a \chi_a\,
\delta(x-x')\delta(t-t'), \quad (a = e,p)
\end{equation*}
where we have dropped the noise corresponding to the viscous term $D_p \partial_x^2 p$ for simplicity. With the identifications above, taking the linear combinations defining $e$ and $p$ in terms of quasiparticles (Eq.~\eqref{eqn:qp-decomp}) reduces Eq.~\eqref{eq:boltzmann_app} to the two-mode fluctuating linear hydrodynamics equation in
Eqs.~\eqref{eq:eps_hydro_app}–\eqref{eq:phi_hydro_app} (after neglecting the viscous/Fickian term). In this form, the quasiparticle picture becomes precise: between stochastic events the quasiparticles propagate ballistically at $\pm c$, while energy-breaking events convert particle-like quasiparticles to hole-like quasiparticles (and reverse direction), and energy-current-breaking events reverse direction without changing the particle/hole type. The correlated or anticorrelated momentum kicks discussed in the main text are simply the statistics of repeated returns generated by these two Poisson processes.

We note that the kinetic description in Eq.~\eqref{eq:boltzmann_app} is not unique~\cite{BENZI1992145}: different microscopic implementations of the Poisson processes (e.g.~redistributing relaxation between the two flip channels or between explicit flips and quasiparticle birth--death noise) lead to the same closed linear equations for $e$ and $p$ with identical transport coefficients and noise correlators. This ``gauge freedom'' of the Boltzmann representation does not affect any hydrodynamic observables.

\subsection{$\gamma_e=\gamma_p$ and the fully conserved case}
A useful simplification occurs on the diagonal line $\gamma_e=\gamma_p$.  Although the quasiparticle labels $(a,s)$ in Eq.~\eqref{eq:boltzmann_app} make the two flip processes look like \emph{reflections}, one can exploit the non-uniqueness of the kinetic representation to choose variables in which no quasiparticle ever reflects: instead, right- and left-movers simply propagate ballistically and decay, with the stationary state maintained
by birth noise.

To see this explicitly, define the signed (particle--hole) densities for each propagation direction,
\begin{equation}
g_s(x,t)\;\equiv\;\sum_{a=\pm} a\, f_{a,s}(x,t) \;=\; f_{+,s}(x,t)-f_{-,s}(x,t),\qquad s=\pm.
\end{equation}
In terms of these fields the hydrodynamic variables are $e = g_{+}+g_{-}$, and $p/c= g_{+}-g_{-}$. Multiplying Eq.~\eqref{eq:boltzmann_app} by $a$ and summing over $a$ gives a closed equation for $g_s$:
\begin{equation}
\partial_t g_s+sc\,\partial_x g_s
=
-\frac{\gamma_e+\gamma_p}{2}g_s
+\frac{\gamma_p-\gamma_e}{2}g_{-s}
+\xi_s(x,t).
\label{eq:gsevol_general}
\end{equation}
where $\xi_s\equiv \sum_a a\,\zeta_{a,s}$ is Gaussian white noise (its covariance is fixed by demanding the correct equal-time susceptibilities for $e$ and $p$). On the diagonal $\gamma_e=\gamma_p\equiv \gamma$, the coupling between opposite directions cancels identically and we obtain
\begin{equation}
\partial_t g_s+sc\,\partial_x g_s=-\gamma g_s+\xi_s(x,t).
\label{eq:gs_decay}
\end{equation}
Thus, in this gauge there is \emph{no backscattering}: right-movers remain right-moving, left-movers remain left-moving, and the only effect of the symmetry-breaking processes is an exponential decay of each chiral density. The stationary state is maintained by the birth--death noise $\xi_s$. This representation makes it particularly transparent why, at $D_p=0$, the convective diffusion constant is independent of the common relaxation rate $\gamma$ for any $\gamma\geq 0$. Since there are no reflections, any quasiparticle crosses the origin at most once. Furthermore, the fluctuation--dissipation condition fixes the birth noise amplitude to be proportional to the decay rate, meaning that the total number of quasiparticles crossing the origin does not depend on $\gamma$. As a result, the time-integrated current autocorrelation $\int_0^\infty dt\,\langle p(t)p(0)\rangle$, and hence $D(\gamma,\gamma)$, is independent of $\gamma$, including the fully conserved case, $\gamma=0$. This once again shows that the weak dissipation limit along the ray $\gamma_e=\gamma_p$ corresponds to the fully conserved case. Reintroducing $D_p>0$ changes $D(\gamma,\gamma)$ at fixed $\gamma \geq 0$, but does not alter the diffusion constant in the weak dissipation limit. 

\section{Stochastic symmetry--broken ideal gas}
\label{app:stochastic_gas}

We define a stochastic gas as a controlled deformation of a one--dimensional ideal gas of impenetrable point particles. This formulation makes explicit which conservation laws are broken and allows a direct demonstration of microscopic reversibility. We begin with a 1D ideal gas of identical particles of unit mass. Particles move ballistically between collisions, $\dot{x}_i = v_i$, and undergo elastic collisions. In this limit particle number, momentum, and energy are conserved, and the equilibrium measure is that of an ideal gas. We now systematically break the conservation laws by modifying the collision rules and adding ``splitting" and ``merging" rules.

First, consider binary ($2\!\to\!2$) collisions. To break energy conservation while retaining momentum and particle number conservation, we conserve the center--of--mass velocity
$V = (v_1 + v_2)/2$ while discarding the relative velocity $u = v_1 - v_2$ and drawing a new relative velocity $u'$ from a Rayleigh distribution
\begin{equation}\label{eqn:rayleigh}
    \rho(u') \propto |u'|\, e^{-u'^2/4T}.
\end{equation}
The outgoing velocities are then $v_{1,2}' = V \pm u'/2$. This construction is closely analogous to the momentum--conserving Lowe--Andersen thermostat~\cite{KoopmanLowe2006}, in which a stochastic bath interaction thermalizes the relative motion at temperature $T$ while preserving the pair center--of--mass velocity (and hence total momentum). Unlike Lowe--Andersen, which applies such updates to nearby pairs (within a cutoff distance) at a Poisson rate independent of their relative speed, here the update is applied at actual collision events whose rate is proportional to the relative speed $|u| = |v_1 - v_2|$, which compensates exactly for the velocity bias in collision rates.

Microscopic detailed balance for these collisions can be checked explicitly. For a collision taking incoming velocities $(v_1,v_2)$ to outgoing velocities $(v_1',v_2')$, detailed balance requires
\begin{equation}
P_{\mathrm{eq}}(v_1,v_2)\,
W\!\left[(v_1,v_2)\to(v_1',v_2')\right]
=
P_{\mathrm{eq}}(v_1',v_2')\,
W\!\left[(v_1',v_2')\to(v_1,v_2)\right],
\label{eq:db_collision}
\end{equation}
where $P_{\mathrm{eq}}$ is the equilibrium measure and $W$ the transition rate. Writing velocities in center--of--mass and relative coordinates, the Maxwell equilibrium weight factorizes as
\begin{equation}
P_{\mathrm{eq}}(v_1,v_2) \propto e^{-v_1^2/2T}e^{-v_2^2/2T}
= e^{-V^2/T}\,e^{-u^2/4T},
\end{equation}
and similarly for the outgoing velocities with $u$ replaced by $u'$. The transition rate for the forward process is given by $W\!\left[(v_1,v_2)\to(v_1',v_2')\right]\propto|u|\rho(u')$, while the reverse process is given by $W\!\left[(v_1',v_2')\to(v_1,v_2)\right]\propto|u'|\rho(u)$. Substituting these expressions into Eq.~\eqref{eq:db_collision}, both sides of the detailed balance condition reduce to $|u| |u'|\, e^{-u'^2/4T-u^2/4T}$. 


To break particle--number conservation while preserving momentum, we
introduce splitting and merging events. Independently, each particle may
split at a constant rate $r_{\mathrm{split}}$, producing two children with
velocities
\begin{equation}
    v_{1,2}=\frac{v}{2}\pm\frac{u}{2},
\end{equation}
where $u$ is drawn from the Rayleigh distribution
Eq.~\eqref{eqn:rayleigh}. Conversely, when two particles collide, they
may merge into a single particle carrying their total momentum
$v=v_1+v_2$, with probability
\begin{equation}
    p_{\mathrm{merge}}(v)=p_0 e^{-v^2/4T}.
\end{equation}
The velocity dependence of the merging probability is required for detailed balance. (Note that changing fluid mass, in this case particle number, but not fluid momentum necessarily breaks Galilean invariance, so that velocity dependence of these number breaking events is expected.) Writing
$u=v_1-v_2$, so that
$v_1^2+v_2^2=v^2/2+u^2/2$, the forward splitting weight is
\begin{equation}
P_{\mathrm{eq}}(v)\,
W_{\mathrm{split}}(v\to v_1,v_2)
\propto
e^{-v^2/2T}\,
r_{\mathrm{split}} |u|e^{-u^2/4T},
\end{equation}
whereas the reverse merging weight is
\begin{align}
P_{\mathrm{eq}}(v_1,v_2)\,
W_{\mathrm{merge}}((v_1,v_2)\to v)
&\propto
e^{-v_1^2/2T}e^{-v_2^2/2T}
|u|p_0e^{-v^2/4T} \\
&=
p_0 |u|e^{-v^2/2T}e^{-u^2/4T}.
\end{align}
For a Poisson gas of mean density $n_0$, the equilibrium collision
flux gives a merging rate per unit length
$n_0^2p_0\sqrt{T/2\pi}$, whereas splitting occurs at rate
$n_0r_{\rm split}$. Detailed balance therefore fixes
\begin{equation}
    r_{\rm split}
    = n_0 p_0\sqrt{\frac{T}{2\pi}}.
\end{equation}
Thus the ratio $r_{\rm split}/p_0$ fixes the mean density of the
grand--canonical equilibrium ensemble.

Finally, momentum conservation may also be broken by allowing the center--of--mass velocity to flip sign with probability $p_{\mathrm{flip}}$ during $2\!\to\!2$ collisions. Since the Maxwell distribution is even in $V$, this move is self--inverse and satisfies detailed balance trivially. Putting these ingredients together, each elementary stochastic move---Rayleigh relative velocity redraws at collisions, splitting and merging events, and center--of--mass velocity flips---satisfies microscopic detailed balance with respect to the same grand--canonical Gibbs measure,
\begin{equation}
P(\{x_i,v_i\}) \propto \sum_N P_{\mathrm{GC}}(N)\prod_{i=1}^N e^{-v_i^2/2T}.
\end{equation}
As a result, equal--time snapshots of the system have ideal--gas static correlations: particle positions are Poisson distributed and velocities are independently Maxwell distributed at temperature $T$, despite the absence of energy, particle--number, or momentum conservation in the dynamics.

At zero momentum and number relaxation, the gas considered here is Galilean invariant and has a
charge-conjugation symmetry at charge neutrality. Neither property is
essential to the hydrodynamic structure of interest: in the next Appendix
we give an explicit construction without either boost symmetry or PHS
which leads to hydrodynamic equations of the same form. 

Furthermore, the processes relaxing conservation laws considered here are deliberately chosen so that
particle-number and momentum relaxation can be tuned independently. A
generic coupling to an environment will instead relax both simultaneously.
In that case the hydrodynamic equations remain unchanged, but
$\gamma_e$ and $\gamma_p$ are effective relaxation rates determined by
the details of the bath coupling. A generic weak bath therefore corresponds simply to approaching the
origin along some nonuniversal ray in the $(\gamma_e,\gamma_p)$ plane.

Note that we always choose to break energy conservation in the stochastic gas. This choice is motivated by graphene hydrodynamics, where energy and energy current play roles analogous to particle number and particle current in the present model. The construction above is therefore a minimal (and simulable) model for these hydrodynamic slow modes, both near and far from equilibrium. 

\subsection{Extension to charged particles}
\label{app:charged_gas}

We now extend the stochastic gas defined above by endowing each particle with a discrete charge label $q \in \{-1,0,+1\}$ in order to introduce a conserved analogue of electric charge, while preserving microscopic reversibility and the ideal--gas character of the stationary state.

The introduction of charge labels does not modify how velocities are updated in any stochastic event. In particular, for binary collisions, splitting events, merging events, and center--of--mass velocity flips, the post--event velocities are sampled exactly as for the chargeless case above, with no dependence on the charge labels. Since the velocity update rules are unchanged, the arguments establishing detailed balance and invariance of the Maxwell distribution at temperature $T$ carry over here also. The velocity sector of the dynamics is therefore completely unaffected by the presence of charge labels.

In the $2\to 2$ collisions, we randomly redraw the charge labels from the same charge sector with probability $p_{\rm mix}$, otherwise the ordering (in space) of the labels is preserved. For example, if two particles collide with labels $(+1,-1)$, then with probability $p_{\rm mix}$, the outgoing labels are drawn randomly from $\{(0,0),(+1,-1),(-1,+1)\}$. Up to here, the charge sector is entirely a spectator to the particle dynamics. However, the particle dynamics are affected by the charge conservation law at merging and splitting events, which now include additional constraints.

In a splitting event, a parent particle of charge $q$ produces two children with charges $(q_1,q_2)$ satisfying $q_1+q_2=q$. The allowed splittings are
\begin{equation}
+1 \to (+1,0)\ \text{or}\ (0,+1), \qquad
0 \to (0,0),\ (+1,-1),\ (-1,+1), \qquad
-1 \to (-1,0)\ \text{or}\ (0,-1),
\end{equation}
with no splitting events producing $(+1,+1)$ or $(-1,-1)$. Merging events are defined as the exact inverses of these splittings: two particles may merge only if their charges sum to a value in $\{-1,0,+1\}$, and when a merge is allowed the child inherits the total charge $q=q_1+q_2$ and the total momentum.

The assignment probabilities of charge labels in splitting events determine the stationary single--particle charge distribution. We denote this distribution by $\pi(q)$, with $\pi(+1)=\pi(-1)$ imposed by charge neutrality. The equilibrium measure for the charged gas then factorizes as
\begin{equation}
P_{\mathrm{eq}}(\{v_i,q_i\}) \propto \prod_i e^{-v_i^2/2T}\,\pi(q_i).
\end{equation}
Microscopic detailed balance for charge--conserving splitting and merging follows by construction. Writing the local balance condition
\begin{equation}
P_{\mathrm{eq}}(v,q)\,
W_{\mathrm{split}}\!\left((v,q)\to (v_1,q_1),(v_2,q_2)\right)
=
P_{\mathrm{eq}}(v_1,q_1;v_2,q_2)\,
W_{\mathrm{merge}}\!\left((v_1,q_1),(v_2,q_2)\to (v,q)\right),
\end{equation}
the splitting and merging weights take the explicit form $W_{\mathrm{split}}=r_{\mathrm{split}}(q)\rho(u)p_{\mathrm{split}}(q\to q_1,q_2)$, and $W_{\mathrm{merge}}=|u|p_{\mathrm{merge}}(q_1,q_2)$, where $u=v_1-v_2$, $\rho(u)$ is the Rayleigh distribution, $p_{\rm merge}(q_1,q_2)$ is the probability that a merge of particles with labels $q_1,q_2$ occurs at a collision, and $p_{\rm split}(q\to q_1,q_2)$ is the probability of a particle with label $q$ splitting into children with labels $q_1,q_2$ conditioned on a splitting event occurring.

As in the chargeless case, the velocity--dependent factors cancel identically, and verifying detailed balance reduces to a purely charge--sector condition,
\begin{equation}
\pi(q)\,r_{\mathrm{split}}(q)\,
p_{\mathrm{split}}(q\to q_1,q_2)
=
\pi(q_1)\pi(q_2)\,
p_{\mathrm{merge}}(q_1,q_2),
\qquad q_1+q_2=q .
\label{eq:charge_balance}
\end{equation}

The probabilities of charge-label outcomes at splitting events therefore determine the stationary single--particle charge distribution $\pi(q)$. A natural and sufficient choice is to take the conditional splitting probabilities
\begin{equation}
p_{\mathrm{split}}(q\to q_1,q_2)
=
\frac{\pi(q_1)\pi(q_2)}
{\sum_{q_1'+q_2'=q}\pi(q_1')\pi(q_2')},
\qquad q_1+q_2=q,
\end{equation}
with the merging probabilities $p_{\mathrm{merge}}(q_1,q_2)$ chosen accordingly so that Eq.~\eqref{eq:charge_balance} holds. With this choice, every allowed splitting process is paired with a merging process of equal weight. We work at charge neutrality $\pi(+1)=\pi(-1)=(1-\pi(0))/2$ and choose $\pi(0)=1/3$ for convenience, for which the conditional splitting probabilities reduce to $p_{\mathrm{split}}(0\to q_1,q_2)=1/3$ for the three allowed channels and $p_{\mathrm{split}}(\pm1\to q_1,q_2)=1/2$ for the two allowed channels. For disallowed charge configurations, such as $(+1,+1)$ or $(-1,-1)$, both the splitting process and the reverse merging process are forbidden. These channels therefore satisfy detailed balance trivially and do not affect the equilibrium measure.

\subsection{Calibration of symmetry-breaking rates}

The symmetry-breaking rates $\gamma_p$ and $\gamma_e$ entering the hydrodynamic theory are not, in general, equal to the bare microscopic probabilities $p_{\rm flp}$ and $p_{\rm merge}$; the mapping involves $O(1)$ nonuniversal coefficients. Rather than attempting to compute these coefficients microscopically, we calibrate the mapping empirically using the measured diffusion constant $D(p_{\rm flp},p_{\rm merge})$. Operationally, we have access to (i) the diffusion constant $\mathfrak{D}$ in the fully conserved limit (no symmetry breaking), obtained numerically, and (ii) a set of measurements of $D$ along a scan of ratios $p_{\rm merge}/p_{\rm flp}$ in the weak-breaking regime. We then define $\gamma_p \equiv p_{\rm flp}$ and $\gamma_e \equiv a\,p_{\rm merge}$, where the calibration constant $a$ is chosen so that $D(\gamma,\gamma)=\mathfrak{D}$, consistent with the result from linear hydrodynamics (Appendix~II) that the diagonal ray $\gamma_e=\gamma_p$ reproduces the fully conserved diffusion constant. This choice simply fixes the relative units of $\gamma_e$ and $\gamma_p$, ensuring that the line $\gamma_e=\gamma_p$ in our plots corresponds, in the weak dissipation limit, to the charge diffusion constant $\mathfrak{D}$. For the data shown in
Fig.~2 this procedure yields $a \simeq 0.213$. We use this conversion, from $(p_{\rm merge},p_{\rm flip})$ to $(\gamma_e,\gamma_p)$, for all figures shown in the main text.

\section{Convective diffusion beyond particle--hole and boost symmetry}
\label{app:generality}

Here we discuss the broader scope of the convective-diffusion mechanism used in
the main text. Particle--hole symmetry provides a particularly simple
route to a diffusive charge mode coexisting with ballistic sound, but is
not essential. A broad class of examples is provided by equilibrium
multicomponent fluids.

Consider first a two-component particle fluid with separately conserved
species densities $\rho_\pm$. We define
\begin{equation}
    \rho=\rho_++\rho_-,
    \qquad
    n=\rho_+-\rho_-,
\end{equation}
and consider a charge neutral equilibrium state, $n_0=0$, without assuming any
symmetry exchanging the two species. In local equilibrium, the microscopic state of a fluid element is
distributed according to the local Gibbs measure
\begin{equation}
    \mathcal P_{\rm LE}
    \propto
    \exp\left[-\beta\left(H-u\Pi-\mu_+N_+-\mu_-N_-\right)\right],
    \label{eq:local_gibbs_general}
\end{equation}
where $u$ is the thermodynamic variable conjugate to the total momentum
$\Pi$. Importantly, introducing $u$ in Eq.~\eqref{eq:local_gibbs_general}
does not require a boost symmetry.

For a continuum particle fluid, $\Pi=\sum_i p_i$, where $i$ labels each particle. Integrating by parts
with respect to any particle momentum,
\begin{equation}
0=
\int dp_i\,
\frac{\partial}{\partial p_i}
\exp\!\bigg[-\beta\bigg(H-u\Pi-\sum_a\mu_aN_a\bigg)\bigg] \implies \left\langle \frac{\partial H}{\partial p_i}\right\rangle_{\rm LE}
=u.
\end{equation}
Using Hamilton's equation $\dot x_i=\partial H/\partial p_i$, each
species therefore has the same Euler-scale drift velocity $u$, so that
$j_a=\rho_a u$. Consequently
\begin{equation}
    j_\rho=\rho u,
    \qquad
    j_n=n u.
    \label{eq:common_advection}
\end{equation}
No particle--hole or boost symmetry has been assumed.

Equation~\eqref{eq:common_advection} is the essential ingredient for the
convective diffusion discussed in the main text. At charge neutrality,
the Euler-scale charge current vanishes identically whenever $n=0$,
irrespective of fluctuations in the other hydrodynamic fields. To express this current in terms of the hydrodynamic densities, we regard
$u$ as a thermodynamic function of $p,\rho,n,e$ (with $e$ omitted when
energy is rapidly relaxed by a thermal bath). Expanding about a
zero-flow equilibrium state, $p=0$, spatial parity implies that to
linear order
\begin{equation}
    u=\chi_p^{-1}p+\ldots ,
\end{equation}
where $\chi_p=(\partial p/\partial u)_{\rm eq}$ is the momentum
susceptibility.
Thus to quadratic order, the Euler-scale current of the charge mode is $j_n \propto np$. This is precisely the mode coupling structure responsible for convective diffusion.

The same constitutive structure is familiar in fluids with boost
symmetry. For Galilean- and Lorentz-invariant perfect fluids, conserved
species densities are advected with the fluid velocity $j_a = \rho_a u$, while the
Euler-scale energy current takes the form $j_e=(e+P)u$ for pressure $P$~\cite{deBoer2018PerfectFluids,heavyionlecturenotes,KovtunNotes}. More generally, the perfect-fluid constitutive relations retain these
forms even without boost symmetry~\cite{deBoer2018PerfectFluids,deBoer2020NonBoost}. A boost symmetry provides an additional relation  between momentum and fluid velocity,
but is not required for the common advection of conserved scalar densities.

These conclusions also hold in the case in which energy is rapidly relaxed by a local thermal bath, while momentum and the two species numbers remain conserved.
Let the bath act locally on the particle momenta and satisfy detailed
balance, within each fixed-$(\Pi,N_+,N_-)$ sector, with respect to a
reservoir temperature $T_b$:
\begin{equation}
e^{-\beta_b H(X)}W(X\to X')
=
e^{-\beta_b H(X')}W(X'\to X),
\end{equation}
for any two microstates $X$ and $X'$ in the same sector.

Since $\Pi$, $N_+$, and $N_-$ have the same values in $X$ and $X'$,
multiplying both sides by
$e^{\beta_b(u\Pi+\mu_+N_++\mu_-N_-)}$ shows that the bath also preserves
the local Gibbs measure Eq.~\eqref{eq:local_gibbs_general}. Integration-by-parts then once again yields the Euler current relation $j_a = \rho_a u$. Thus rapidly relaxing energy removes it as a hydrodynamic mode without
changing the Euler currents of the remaining conserved scalar densities.

The symmetry-free argument above relies on the equilibrium local-Gibbs
structure. In nonequilibrium systems, such as coupled driven exclusion
processes, the Euler currents need not take the common-advection form
$j_a=\rho_a u$. Consequently, $j_n=0$ at $n=0$ is not automatic and
must instead be enforced by an additional symmetry, such as PHS, or by
microscopic tuning~\cite{Ferrari2013,
Popkov2015,Spohn2015}.

\section{Hydrodynamic recoupling with an additional slow mode}
\label{app:additional_mode}

We now show that the hydrodynamic-recoupling mechanism is unchanged
when an additional scalar hydrodynamic mode is retained, provided it is also weakly relaxed. For simplicity
we assume particle--hole symmetry (PHS) in this section, so that the
charge mode of interest is linearly decoupled from the remaining
hydrodynamic fields at charge neutrality.

We denote the conserved charge density by $n$, the momentum density by
$p$, the energy density by $e$, and an additional PHS-even scalar density $\rho$. In a
multicomponent particle fluid, $\rho$ is naturally the total particle
density. In charge-neutral graphene, the same hydrodynamic structure
arises if $\rho$ is identified with a long-lived carrier-imbalance density.
When electron--hole collisions efficiently exchange momentum, as in the
hydrodynamic Dirac-fluid regime~\cite{PhysRevB.78.085416,Muller2008,
Ponomarenko2024}, electrons and holes share a common local fluid velocity.
Carrier imbalance therefore enters as an additional scalar hydrodynamic mode,
rather than as a second momentum-carrying fluid. Its approximate conservation
and associated relaxation scale in graphene have been discussed extensively,
for example in Ref.~\cite{PhysRevB.79.085415}.

We expand around a homogeneous zero-flow, charge-neutral equilibrium
state, $e=e_0+\delta e$, $\rho=\rho_0+\delta\rho$, with $n_0=p_0=0$. For the locally equilibrating fluids considered above, the Euler currents of the scalar densities take the usual
perfect-fluid form~\cite{deBoer2018PerfectFluids}
\begin{equation}
    j_e=(e+P)u,\qquad
    j_\rho=\rho u, \qquad j_n = n u,
    \label{eq:perfect-fluid-currents}
\end{equation}
where $u$ is the local fluid velocity. To linear order about the
zero-flow state, $u=\chi_p^{-1}p+\ldots$. The momentum current is the stress tensor. In local equilibrium its value in
the local rest frame is the pressure, so that to linear order $\delta j_p=\delta P$. At PHS, the pressure has no term linear in $n$, and spatial parity symmetry
forbids a term linear in $p$. Thus
\begin{equation}
    \delta P=P_e\,\delta e+P_\rho\,\delta\rho,
    \qquad
    P_e\equiv
    \left.\frac{\partial P}{\partial e}\right|_0,
    \qquad
    P_\rho\equiv
    \left.\frac{\partial P}{\partial\rho}\right|_0 .
\end{equation}
The linear Euler equations in the PHS-even sector are therefore
\begin{align}
    \partial_t\delta e
    +\frac{e_0+P_0}{\chi_p}\partial_xp =0,
    \qquad 
    \partial_t\delta\rho
    +\frac{\rho_0}{\chi_p}\partial_xp =0,
    \qquad
    \partial_t p
    +P_e\partial_x\delta e
    +P_\rho\partial_x\delta\rho &=0.
    \label{eq:four_mode_euler_explicit}
\end{align}
It is convenient to collect the scalar fluctuations into a vector $\boldsymbol{\phi} = (\delta e,\delta\rho)^T$, so that
\begin{equation}
    \partial_t\boldsymbol{\phi}
    +\mathbf a\,\partial_xp=0,
    \qquad
    \partial_t p
    +\mathbf b^{T}\partial_x\boldsymbol{\phi}=0,
    \label{eq:four_mode_euler}
\end{equation}
where $\mathbf a= \chi_p^{-1} (e_0+P_0, \rho_0)^T$, and $\mathbf b=(P_e, P_\rho)^T$.
This three-field sector contains two sound modes with speed of sound $c$ given by $c^2=\mathbf b^T\mathbf a$, and one zero-velocity mode. Equivalently, one linear combination
of $\delta e$ and $\delta\rho$ participates in sound, while an independent linear combination is the usual heat mode. We will not need to identify these combinations explicitly.

\subsection{Comment on the heat mode}

It is useful to ask why an ordinary fluid with only particle-number,
momentum, and energy conservation does not already realize the same
mechanism, with the zero-velocity heat mode playing the role of charge.
Such a fluid has precisely the same linear mode counting as the theory
considered in the main text: two sound modes with velocities $\pm c$ and one
zero-velocity mode.  The distinction appears at nonlinear order and
follows directly from the different perfect-fluid currents in
Eq.~\eqref{eq:perfect-fluid-currents}.

To see this, first define $w_0=e_0+P_0$, and denote by $h$ the zero-velocity normal mode, given by
\begin{equation}
    h = \delta e-\frac{w_0}{\rho_0}\delta\rho .
\end{equation}
Indeed, Eq.~\eqref{eq:four_mode_euler_explicit} gives $\partial_t h=0$ at
linear Euler order.  Since $h$ is a linear combination of conserved
densities, its Euler current is
\begin{align}
    j_h
    = j_e-\frac{w_0}{\rho_0}j_\rho
    = \left[e+P-\frac{w_0}{\rho_0}\rho\right]u 
    = \left(h+\delta P\right)u+O(\delta^3).
\end{align}
Thus, unlike the charge current, the heat current does not vanish when
the heat-mode density is set to zero:
\begin{equation}
    \left.j_h\right|_{h=0}
    = \delta P\,u+O(\delta^3).
\end{equation}
A pure sound fluctuation has both $\delta P\neq0$ and $u\neq0$, so this
term produces the familiar diagonal mode coupling
\begin{equation}
    j_h \sim \phi_+^2-\phi_-^2 ,
\end{equation}
where $\phi_\pm$ denote the right- and left-moving sound modes.  The
zero-velocity heat mode is therefore generically superdiffusive already in
the fully conserved theory, leading to the familiar $z=5/3$ heat-mode
scaling of nonlinear fluctuating hydrodynamics
\cite{SpohnNLFH,Delacretaz2020}.

This should be contrasted with the charge (or species-imbalance) mode.
Local equilibrium gives $j_n=nu$,
so that $j_n$ vanishes identically whenever
$n=0$, irrespective of the sound fluctuations.  Consequently the
diagonal coupling $j_n\sim\phi_+^2-\phi_-^2$ is absent, and the charge
mode can remain diffusive in the fully conserved theory.

\subsection{Weak relaxation of momentum, particle number, and energy}

We now weakly relax momentum and both $e$ and $\rho$, while conserving $n$. A generic bath need not relax $e$ and $\rho$
independently, so the most general linear relaxation in the scalar sector
is described by a matrix $\Gamma$:
\begin{align}
    \partial_t\boldsymbol{\phi}
    +\mathbf a\,\partial_xp
    =-\Gamma\boldsymbol{\phi}
    +\boldsymbol{\zeta}_{\phi},
    \qquad
    \partial_t p
    +\mathbf b^T\partial_x\boldsymbol{\phi}
    =-\gamma_p p+\zeta_p .
    \label{eq:four_mode_relaxed}
\end{align}
The reservoirs are taken to be matched to the reference equilibrium
state (equivalently, one may first equilibrate the system with the reservoirs before varying their coupling strength), so that the reference state remains stationary under the weak relaxation.
We omit intrinsic diffusive and viscous terms in this sector for
simplicity. The calculation in Sec.~\ref{app:linear_hydro}, in which these terms are
retained, shows that they give only subleading corrections to the
convective contribution to the charge diffusion constant in the
weak-relaxation limit.

As in the three mode calculation, the convective contribution to charge
diffusion is determined entirely by the momentum fluctuations sampled by
the charge mode. In particular, Eq.~\eqref{eqn:DX} gives
\begin{equation}
    D_{\rm conv}
    =
    \chi_p^{-2} \int\frac{dk}{4\pi}\,S_{pp}(k,0).
    \label{eq:Dconv_four_mode}
\end{equation}
As in the three mode case of Sec.~\ref{app:linear_hydro}, we now determine the leading weak relaxation limit of the zero-frequency momentum
spectral function $S_{pp}(k,0)$.

To study the joint weak-relaxation limit, we introduce an overall
relaxation scale $\gamma$ and write $\Gamma=\gamma\widehat{\Gamma}$, $\gamma_p=\gamma\widehat{\gamma}_p$, taking $\gamma\to0$ at fixed $\widehat{\Gamma}$ and
$\widehat{\gamma}_p$. Fourier transforming the linearized equations in space and time, we obtain
\begin{equation}
    \left[-i\omega+\mathsf M(k)\right]
    \boldsymbol{\Psi}(k,\omega)
    =
    \boldsymbol{\xi}(k,\omega),
\end{equation}
where
\begin{equation}
    \boldsymbol{\Psi}(k,\omega)
    =
    \begin{pmatrix}
        \boldsymbol{\phi}(k,\omega)\\
        p(k,\omega)
    \end{pmatrix},
    \qquad
    \boldsymbol{\xi}(k,\omega)
    =
    \begin{pmatrix}
        \boldsymbol{\zeta}_{\phi}(k,\omega)\\
        \zeta_p(k,\omega)
    \end{pmatrix},
    \qquad 
    \mathsf M(k)=
    \begin{pmatrix}
        \gamma\widehat{\Gamma} & ik\,\mathbf a\\
        ik\,\mathbf b^T & \gamma\widehat{\gamma}_p
    \end{pmatrix}.
\end{equation}
Since Eq.~\eqref{eq:Dconv_four_mode} requires only zero frequency, we require only 
\begin{equation}
    \boldsymbol{\Psi}(k,0)=\mathsf M^{-1}(k)\boldsymbol{\xi}(k,0).
    \label{eq:four_mode_zero_frequency_solution}
\end{equation}

The relevant wavevectors are those for which the Euler terms,
proportional to $k$, compete with the weak relaxation terms,
proportional to $\gamma$. Hence we write $k=\gamma\kappa$.
At these wavevectors, $\mathsf M(\gamma\kappa)=\gamma\widehat{\mathsf M}(\kappa)$ and therefore $\mathsf M^{-1}(\gamma\kappa) =\gamma^{-1}\widehat{\mathsf M}^{-1}(\kappa)$.

By the fluctuation--dissipation theorem, the covariance of the
bath-induced noise is likewise proportional to $\gamma$. Writing
\begin{equation}
    \langle
    \boldsymbol{\xi}(k,\omega)
    \boldsymbol{\xi}^\dagger(k',\omega')
    \rangle
    \propto
    \gamma\widehat{\mathsf Q}\,
    \delta(k-k')\delta(\omega-\omega'),
\end{equation}
we define
\begin{equation}
    \widehat S_{pp}(\kappa)
    \equiv
    \left[
        \widehat{\mathsf M}^{-1}(\kappa)
        \widehat{\mathsf Q}
        \widehat{\mathsf M}^{-\dagger}(\kappa)
    \right]_{pp}.
\end{equation}
It follows that
\begin{equation}
    S_{pp}(\gamma\kappa,0)
    =
    \gamma^{-1}\widehat S_{pp}(\kappa).
\end{equation}

Substitution into Eq.~\eqref{eq:Dconv_four_mode}, together with
$dk=\gamma\,d\kappa$, therefore gives
\begin{equation}
    D_{\rm conv}
    \xrightarrow[]{\gamma\to0}
    \chi_p^{-2}\int\frac{d\kappa}{4\pi}\,
    \widehat S_{pp}(\kappa;
        \widehat{\Gamma},\widehat{\gamma}_p).
    \label{eq:Dconv_four_mode_limit}
\end{equation}
Thus the convective diffusion constant approaches a finite value which
depends on the ratios of relaxation rates, $\hat \gamma_p$ and $\hat \Gamma$. The
hydrodynamic-recoupling singularity therefore persists in the presence
of the additional scalar mode.

For carrier imbalance, this result can be understood in terms of its
relaxation rate $\gamma_I$. If $\gamma_I$ is parametrically larger than the
weak energy- and momentum-relaxation rates, imbalance ceases to be a
hydrodynamic mode on the scales of interest and the three-mode theory of the
main text is recovered. If instead $\gamma_I$ is of the same parametric order
as $\gamma_e$ and $\gamma_p$, it must be retained as above, and the
weak-relaxation diffusion constant depends on the additional ratio
$\gamma_I/\gamma_p$ as well as $\gamma_e/\gamma_p$.

The latter regime is also physically natural. A weak coupling to an
environment that exchanges energy, and does not separately protect carrier
imbalance, will generally relax imbalance as well. As this coupling is reduced, one naturally expects $\gamma_e$ and $\gamma_I$
to vanish together with a finite nonuniversal ratio. Only if imbalance is exactly conserved, or parametrically
longer lived than the other weakly relaxed modes, does it remain an additional
infrared hydrodynamic degree of freedom that can modify the asymptotic
universality class.

\end{document}